\pdfoutput=1

\documentclass[11pt]{article}

\usepackage[final]{acl}

\usepackage{times}
\usepackage{latexsym}

\usepackage[T1]{fontenc}

\usepackage[utf8]{inputenc}

\usepackage{microtype}

\usepackage{inconsolata}

\usepackage{multirow}
\usepackage[T5]{fontenc}

\usepackage{listings}
\usepackage{tabularx}
\usepackage{ltablex}
\usepackage{longtable}
\usepackage{graphicx}
\usepackage{colortbl}

\usepackage{pythonhighlight}

\usepackage{supertabular}
\usepackage{xltabular}

\usepackage[toc,page,header]{appendix}
\usepackage{minitoc}
\usepackage{tcolorbox}
\usepackage{cleveref}
\usepackage{todonotes}

\title{Medical Spoken Named Entity Recognition}

\author{Khai Le-Duc$^{1,2}$, David Thulke$^{4,5}$, Hung-Phong Tran$^{3}$, Long Vo-Dang$^{7}$, \\ {\bf Khai-Nguyen Nguyen$^{8}$, Truong-Son Hy$^{6}$, Ralf Schlüter$^{4,5}$}\\
$^1$University of Toronto, Canada
$^2$University Health Network, Canada \\
$^3$Hanoi University of Science and Technology, Vietnam\\
$^4$Machine Learning and Human Language Technology Group,\\ RWTH Aachen University, Germany\\
$^5$AppTek GmbH, Germany
$^6$University of Alabama at Birmingham, United States \\
$^7$University of Cincinnati, United States
$^8$College of William and Mary, United States
\\
  \texttt{duckhai.le@mail.utoronto.ca} \\
  \texttt{thy@uab.edu, \{thulke,schlueter\}@hltpr.rwth-aachen.de} 
  \\}

\begin{document}
\maketitle

\thispagestyle{plain}
\pagestyle{plain}

\begin{abstract}
Spoken Named Entity Recognition (NER) aims to extract named entities from speech and categorise them into types like person, location, organization, etc.  
In this work, we present \textit{VietMed-NER} - the first spoken NER dataset in the medical domain.
To our knowledge, our Vietnamese real-world dataset is the largest spoken NER dataset in the world regarding the number of entity types, featuring 18 distinct types.
Furthermore, we present baseline results using various state-of-the-art pre-trained models: encoder-only and sequence-to-sequence; and conduct quantitative and qualitative error analysis.
We found that pre-trained multilingual models generally outperform monolingual models on reference text and ASR output and encoders outperform sequence-to-sequence models in NER tasks.
By translating the transcripts, the dataset can also be utilised for text NER in the medical domain in other languages than Vietnamese.
All code, data and models are publicly available\footnote{\label{footnote:repo}\url{https://github.com/leduckhai/MultiMed/tree/master/VietMed-NER}}.
\end{abstract}




\section{Introduction}
Named Entity Recognition (NER) targets extracting named entities (NE) from text and categorizing them into types like person, location, organization, etc. 
Initially studied in written language, recent attention has turned to study spoken NER  \cite{cohn2019audio_deID,shon2022slue_dataset}, which aims to extract semantic information from speech.
However, spoken NER has limited literature compared to NER on written text data \cite{yadav2020end2end_ner}.

Spoken NER is particularly challenging, firstly due to the impact of word segmentation on results. 
The medical vocabulary poses difficulties with numerous confused monosyllabic and polysyllabic words.
For instance, the word "đường" alone could denote "sugar" (chemical), "street" (location), or be part of a compound word like "đường tiêu hóa" - "gastrointestinal" (anatomy).
This confusion has also been reported in Chinese spoken NER by \citet{chen2022AISHELL_NER_dataset}.
Further, data quality control and annotation consistency have been problematic, with some entities tagged in one sentence but not in others, and full NEs inconsistently tagged as multiple sub-NEs \cite{huyen2016vlsp_dataset_NER, nguyen2018vlsp_dataset_NER, nguyen2020improving_NER_vietnamese, PhoNER_COVID19}.
Finally, obtaining accurate medical NER from natural speech is challenging due to the lack of punctuation \cite{ertopccu2017new}, speech disfluencies \cite{kim2000rule}, context, and the complexity of medical terms.


As for the medical domain, to the best of our knowledge, there is no dataset available for medical spoken NER.
The only related work we found, \cite{cohn2019audio_deID}, published a NER evaluation benchmark using an English general-domain conversational dataset, Switchboard \cite{godfrey1992switchboard_dataset} and Fisher \cite{cieri-etal-2004-fisher}, for the task of audio de-identification specifically targeting Personal Health Identifiers.

To address this gap, we introduce \textit{VietMed-NER}, a medical spoken NER dataset built on the real-world medical Automatic Speech Recognition (ASR) dataset \textit{VietMed} \cite{vietmed_dataset}, featuring 18 medically-defined entity types.
In the era of the advanced in-context learning capabilities of Large Language Models (LLMs) and human-level text-to-speech technologies, the dataset, with entity positional labels maintained during translation, is applicable not only to Vietnamese but also to other languages (see \Cref{sec:possible_applications}). This enables various real-world applications, including:
search engines \cite{rud2011piggyback_searchengine_ner}, 
content classification for news providers \cite{kumaran2004textclassify_ner}, medical ASR error correction \cite{mani2020ASR_errorcorrection_medical}, audio de-identification \cite{cohn2019audio_deID} and content recommendation systems \cite{koperski2017content_recommendation_ner}.

Our contributions are as follows:
\begin{itemize}
    \item We present \textit{VietMed-NER} - the first publicly-available medical spoken NER dataset. 
    \item We present baselines on several state-of-the-art pre-trained models
    \item We conduct quantitative and qualitative error analysis for medical spoken NER in Vietnamese
\end{itemize}

All code, data and models are published online\footnotemark[1].


\begin{table*}
\centering
\resizebox{\textwidth}{!}{%
\begin{tabular}{|ll|rr|rr|rr|rr|}
\hline
\multicolumn{1}{|c|}{\multirow{2}{*}{}} & \multicolumn{1}{c|}{\multirow{2}{*}{\textbf{Definition}}} & \multicolumn{2}{c|}{\textbf{Train}} & \multicolumn{2}{c|}{\textbf{Dev}} & \multicolumn{2}{c|}{\textbf{Test}} & \multicolumn{2}{c|}{\textbf{All}} \\ \cline{3-10} 
\multicolumn{1}{|c|}{} & \multicolumn{1}{c|}{} & \multicolumn{1}{c|}{Total} & \multicolumn{1}{c|}{Uni.} & \multicolumn{1}{c|}{Total} & \multicolumn{1}{c|}{Uni.} & \multicolumn{1}{c|}{Total} & \multicolumn{1}{c|}{Uni.} & \multicolumn{1}{c|}{Total} & \multicolumn{1}{c|}{Uni.} \\ \hline
\multicolumn{1}{|l|}{AGE} & Age of a person & \multicolumn{1}{r|}{447} & 43 & \multicolumn{1}{r|}{108} & 25 & \multicolumn{1}{r|}{611} & 83 & \multicolumn{1}{r|}{1166} & 151 \\ \hline
\multicolumn{1}{|l|}{GENDER} & Gender of a person & \multicolumn{1}{r|}{202} & 30 & \multicolumn{1}{r|}{46} & 15 & \multicolumn{1}{r|}{451} & 33 & \multicolumn{1}{r|}{699} & 78 \\ \hline
\multicolumn{1}{|l|}{JOB} & Job of a person & \multicolumn{1}{r|}{543} & 32 & \multicolumn{1}{r|}{133} & 16 & \multicolumn{1}{r|}{562} & 43 & \multicolumn{1}{r|}{1238} & 91 \\ \hline
\multicolumn{1}{|l|}{LOCATION} & Locations and places & \multicolumn{1}{r|}{284} & 66 & \multicolumn{1}{r|}{76} & 31 & \multicolumn{1}{r|}{317} & 75 & \multicolumn{1}{r|}{677} & 172 \\ \hline
\multicolumn{1}{|l|}{ORGANIZATION} & Organizations & \multicolumn{1}{r|}{19} & 14 & \multicolumn{1}{r|}{2} & 2 & \multicolumn{1}{r|}{58} & 23 & \multicolumn{1}{r|}{79} & 39 \\ \hline
\multicolumn{1}{|l|}{DISEASESYMTOM} & Symptoms and diseases & \multicolumn{1}{r|}{2699} & 518 & \multicolumn{1}{r|}{683} & 209 & \multicolumn{1}{r|}{1334} & 357 & \multicolumn{1}{r|}{4716} & 1084 \\
\hline
\end{tabular}%
}
\caption{Entity definition and its statistics in our dataset. "Uni." means the number of unique entities.}
\label{data_stats}
\end{table*}

\section{Related Works}
Traditionally, spoken NER has been done using a pipeline methodology, also known as cascaded approach, starting with an ASR stage, followed by NER applied to the generated transcriptions \cite{jannet2017investigating, benaicha2024leveraging}. Another variant of the cascaded approach involves embedding specific entity expressions into the lexicon, thereby improving the language model's ability to accurately recognize these expressions \cite{hatmi2013incorporating}.

Besides, end-to-end NER has recently garnered some attention within the research community. This approach seeks to optimize ASR and NER processes simultaneously, offering a potentially more efficient alternative to traditional pipeline methods by harnessing the ability of trainable acoustic features. However, its accuracy advantage over the cascaded approach remains a subject of debate, and the end-to-end training setup introduces additional complexity \cite{tomashenko2019recent, yadav2020end2end_ner}.

\section{Data}
\subsection{Data Collection}
\label{sec:Data_Collection}

We chose the \textit{VietMed} dataset \cite{vietmed_dataset}, the world's largest publicly available medical ASR dataset, for annotating NEs. 

The original dataset is in Vietnamese.
We annotate the Vietnamese version with the methodology described in \Cref{sec:annotation_process} and automatically translate the transcripts to English together with transferring the NE annotation.

\subsection{Annotation Process}
\label{sec:annotation_process}
The annotation of medical NEs from real-world speech is challenging because of the missing punctuation, special characters and capitalized words in ASR transcripts, disfluencies and required medical knowledge.
Entirely manual annotation of NEs like in VLSP dataset \cite{huyen2016vlsp_dataset_NER, nguyen2018vlsp_dataset_NER, nguyen2020improving_NER_vietnamese} and PhoNER\_COVID19 \cite{PhoNER_COVID19} requires a large number of working hours, not to mention the difficulties in quality control and inconsistency as we found in their corpora.
These inconsistencies include: i) Some entities tagged in one sentence are not tagged in another sentence, and ii) Full NEs are inconsistently tagged as multiple sub-NEs.
The best approach to tag nested NEs is the subject of ongoing debate \cite{muis-lu-2017-labeling,li-etal-2021-span}. For simplicity, higher consistency and to reduce the annotation effort, we only annotate the largest and outermost full entity span.

Moreover, using fine-tuned models for pre-tagging doesn't apply to specific medical entity types. 
Similarly, using prompt engineering with large language models like GPT-4 for pre-tagging did not achieve acceptable accuracy.
Training a seed model with a gazetteer list requires initial training time, subsequent repetitive training schedules, and may prove unreliable due to its statistical reliance on a small amount of low-resource data \cite{kozareva2006NER_gazetteer_list}.


To tackle these problems, we conduct a human-machine annotation approach, as described below:
\begin{enumerate}
    \item Annotate and categorize a set of initial entities, then add them to a gazetteer list.
    \item Sort entities by character length from highest to lowest, to distinguish between sub-NEs and full NEs, ensuring full NEs are mapped before sub-NEs. 
    Time complexity = $O(k\cdot\log(k))$ where $k$ is the number of NEs.
    For example, "tooth pain" should be mapped before "pain".
    \item Automatically map entities from the gazetteer list to the transcript. 
    Time complexity = $O(m\cdot n)$, where $m$ is the number of NEs in gazeeter list, $n$ is the number of sentences.
    Pseudo code:
\begin{python}
for NE in gazetteer_list:
    for sen in sentences:
        if NE in sen:
            annotate(NE, sen)
\end{python}
    \item Annotators review each sentence to include correctly labeled NEs and ignore mislabeled NEs 
    \item Annotators add new NEs not in the gazetteer list during manual annotation. 
    Steps 2 and 3 generate pre-tagged labels in the next sentences. 
    Annotators repeat Steps 4 and 5 until the entire corpus is annotated.
\end{enumerate}

We experience faster annotation by allowing annotators to foresee possible NEs in upcoming utterances based on previously annotated ones. 
Annotators can accept or reject these suggestions, saving time with correct suggestions and easily ignoring incorrect ones. 
Unlike training a seed model with a gazetteer, which requires initial training time and may be unreliable for low-to-mid resource languages, our method avoids these issues and eliminates the need to correct incorrect NEs.

\subsection{Data Quality Control}
We created initial annotation guidelines (see \Cref{sec:annotation_guidelines}) and began annotating the corpus. 
Two developers, one with a medical background, independently annotated the corpus. 
Then, we held a discussion session to resolve conflicts, address complex cases, and refine the guidelines.
Two other developers perform quality control using the guidelines and the annotated corpus. 
We consistently revisited each sentence in the entire corpus multiple times.
This data quality control process is inspired by \citet{tran2021bartpho}. 

\subsection{Data Splitting}
Most NER datasets have a very small number of entities in their test sets compared to train and dev set \cite{huyen2016vlsp_dataset_NER, PhoNER_COVID19, chen2022AISHELL_NER_dataset}.
However, we want to leverage the capabilities of large pre-trained models which are trained on vast amounts of unlabeled text data, resulting in good representations. 
Therefore, we focus on creating a large test set to obtain more statistically significant evaluation results and keep the training set relatively small in comparison.

\subsection{Data Statistics}
\Cref{data_stats} shows the statistics of our dataset.
Our \textit{VietMed-NER} contains 18 entity types across 9000 sentences, split into train-dev-test as 8-2-6 hours. To the best of our knowledge, compared to all other public spoken NER datasets, ours has the largest number of entity types.


\section{Experimental Setups}
We employ the cascaded (two-stage) pipeline for spoken NER: A hybrid ASR model transcribes audio into text and then the transcribed text is fed into a text NER model.

\subsection{Evaluation Metrics}
We employed the F1 score metric as it is commonly used for spoken NER \cite{shon2022slue_dataset,benaicha2024leveraging}, which evaluates an unordered list of NE phrases and tag pairs predicted for each sentence. We used 3 toolkits for a more comprehensive comparison, as described in  \Cref{sec:details_experimental_setups}.

\subsection{ASR Models}

We employed two baseline models fine-tuned for ASR on \textit{VietMed} published by \citet{vietmed_dataset}: an acoustic monolingual pre-trained w2v2-Viet and an acoustic multilingual pre-trained XLSR-53-Viet model.
w2v2-Viet model was pre-trained from scratch on 1204h of Vietnamese data. For the XLSR-53-Viet model, continued pre-training on 1204h of Vietnamese starting with XLSR-53 \cite{conneau21_XLSR53} was performed. Both have the same number of parameters (118M) and were fine-tuned on the same training set.
Their WERs on the test set are 29.0\% and 28.8\% respectively.


\subsection{NER Models}
\begin{table}
\centering
\begin{tabular}{|l|r|r|}
\hline
\textbf{Model}            & \multicolumn{1}{l|}{\textbf{\#Params}} & \multicolumn{1}{l|}{\textbf{\#Data}} \\ \hline
PhoBERT\_base    & 135M                          & \multirow{2}{*}{20GB}       \\ \cline{1-2}
PhoBERT\_large   & 370M                          &                             \\ \hline
PhoBERT\_base-v2 & 135M                          & 140GB                       \\ \hline
ViDeBERTa\_base  & 86M                           & 298GB                       \\ \hline
XLM-R\_base      & 270M                          & 2.5TB                       \\ \hline
XLM-R\_large     & 550M                          & 2.5TB                       \\ \hline
mBART-50         & 611M                          & 3.9TB                       \\ \hline
ViT5\_base       & 310M                          & 888GB                       \\ \hline
BARTpho  & 396M                          & 20GB                        \\ \hline
\end{tabular}
\caption{Statistics of state-of-the-art pre-trained language models which we used for NER task.}
\label{NER_model_stats}
\end{table}
\Cref{NER_model_stats} shows the statistics of various pre-trained monolingual and multilingual models we consider to fine-tune on our dataset. To our knowledge, these are the best pre-trained models that achieved state-of-the-art results on various downstream tasks in the Vietnamese language, including NER.

\textbf{Monolingual encoder models}: {PhoBERT}\_{base}, {PhoBERT}\_{large}, {PhoBERT}\_{base-v2}  \cite{phobert}, {ViDeBERTa}\_{base} \cite{tran2023videberta}. 

\textbf{Monolingual sequence-to-sequence (seq2seq) models}: BARTpho \cite{tran2021bartpho}, ViT5 \cite{phan2022vit5}, 

\textbf{Multilingual encoder models}: {XML-R}\_{base}, {XML-R}\_{large} \cite{conneau2020xlmr}. 

\textbf{Multilingual seq2seq models}: mBART-50 \cite{tang2020multilingual}.

\subsubsection{Seq2seq Training for NER Task}
Following the approach proposed by \citet{phan2021scifive} and later adopted by ViT5 \cite{phan2022vit5}, we formulated the sequence tagging task as a sequence-to-sequence task by training the models to generate tags of labels before and after an entity token. 
In cases where the models fail to follow the mentioned format for an entity token, we use an “exception” tag, which will be later ignored during metric calculation, as the label. 

\subsubsection{Training Hyperparameters}
\label{sec:training_hyperparams}
We used HuggingFace Transformers \cite{wolf2019huggingface} for fine-tuning pre-trained models for the NER task.
Vietnamese input sentences can be represented in either syllable or word level as described by \citet{PhoNER_COVID19}. However, we only employed word-level settings to train NER models.
All our NER experiments were done by using the default hyperparameters by HuggingFace.

The default hyperparameters are as follows:
Learning rate of 2e-5, linear learning rate scheduler, training batch size of 64, 50 training epochs, weight decay of 0.01, AdamW optimizer \cite{loshchilov2018AdamW_SGDW}, Beta1 of 0.9, Beta2 of 0.999, and epsilon of 1e-8.

\section{Experimental Results}
\begin{table}
\centering
\begin{tabular}{|l|c|c|c|}
\hline
\textbf{NER Model} & \textbf{Prec.} & \textbf{Rec.} & \textbf{F1} \\ \hline
BARTpho         & 0.64 & 0.73 & 0.68 \\
mBART-50        & 0.64 & 0.66 & 0.65 \\
PhoBERT\_base   & 0.67 & 0.78 & 0.72 \\
PhoBERT\_base-v2& 0.68 & 0.79 & 0.74 \\
PhoBERT\_large  & 0.69 & 0.77 & 0.73 \\
ViDeBERTa\_base & 0.50 & 0.41 & 0.45 \\
ViT5\_base      & 0.64 & 0.74 & 0.69 \\
XLM-R\_base     & 0.64 & 0.73 & 0.69 \\
XLM-R\_large    & 0.71 & 0.77 & 0.74 \\ \hline
\end{tabular}
\caption{NER results on reference text of test set. The metrics shown are Precision, Recall, and overall micro F1 score.
Results by entity types are shown in Tables \ref{appx_NERresults_AGE}-\ref{appx_NERresults_OverallMicro} in the Appendix.}
\label{NER_results}
\end{table}
\begin{table}[]
\resizebox{\columnwidth}{!}{%
\begin{tabular}{|l|l|c|c|c|}
\hline
\multicolumn{1}{|c|}{\textbf{NER}} & \multicolumn{1}{c|}{\textbf{ASR}} & \multicolumn{1}{c|}{\textbf{Prec.}} & \multicolumn{1}{c|}{\textbf{Rec.}} & \multicolumn{1}{c|}{\textbf{F1}} \\ \hline


\small{ViDeBERTa\_base} & \small{XLSR-53-Viet} & 0.45 & 0.34 & 0.39 \\ \cline{2-5} 
 & \small{w2v2-Viet} & 0.45 & 0.34 & 0.39 \\ \cline{1-5} 
ViT5\_base & \small{XLSR-53-Viet} & 0.52 & 0.46 & 0.48 \\ \cline{2-5} 
 & \small{w2v2-Viet} & 0.53 & 0.46 & 0.49 \\
\hline

\end{tabular}
}
\caption{NER results on ASR output of test set for different NER and ASR models. Metrics shown are Precision, Recall, and overall micro F1 score.
Results by entity types are shown in Tables \ref{appx_spk_NERresults_AGE}-\ref{appx_spk_NERresults_OverallMicro} in the Appendix.}
\label{SpokenNER_results}
\end{table}

Table \ref{NER_results} and \ref{SpokenNER_results} show results of NER using various pre-trained models. We observe that there was a performance drop in all models when evaluated on ASR transcripts, as expected due to the noisy nature of ASR output. 

\textbf{1. Pre-trained multilingual models outperformed monolingual models, if multilingual models overcome the capacity dilution}: The pre-trained monolingual model {PhoBERT}\_{base-v2} outperformed other monolingual models, at 0.74 of F1 score on reference text, and 0.57 on ASR output. Despite having fewer parameters than {PhoBERT}\_{large}, it performed similarly, likely due to more pre-training data. The pre-trained multilingual model {XLM-R}\_{large} achieved the best performance with an F1 score of 0.74 on reference text and 0.58 on ASR output, while {XLM-R}\_{base} performed worse than {PhoBERT}\_{base-v2}. This gap is explained by the larger pre-training data (2.5TB multilingual data for XLM-R vs. 140GB monolingual data for {PhoBERT}\_{base-v2}). Our results with {PhoBERT}\_{base-v2} and {XLM-R}\_{large} confirmed that pre-trained multilingual representations improve performance on medical spoken NER tasks, similar to other language-specific downstream tasks by \citet{conneau2020xlmr, liu2020multilingual}. However, multilingual models may face a \textit{Transfer-dilution Trade-off} \cite{conneau2020xlmr}, where they lack the capacity to learn effective multilingual representations. In other words, for a fixed sized model, the per-language performance decreases as we increase the number of languages \cite{gurgurov2024multilingual}. To address this trade-off, multilingual models should possess sufficient capacity, necessitating an adequately large model size \cite{chen2024efficient}.
This is evident in the performance comparison between {PhoBERT}\_{base-v2} and {XLM-R}\_{base}, as seen in other language-specific downstream tasks by \citet{conneau2020xlmr, arivazhagan2019massively}.

\textbf{2. Encoder-based models outperform seq2seq models}: The best seq2seq model, BARTpho, achieved F1 scores of 0.68 on reference text and 0.53 on ASR output. Encoders generally performed better than seq2seq models, possibly because seq2seq's generative nature is less suited for classification tasks like NER.

\paragraph{3. Multi-lingual pre-training of the acoustic model does not affect cascaded NER performance} As expected by the similar WERs for the acoustic pre-trained monolingual model w2v2-Viet and the multilingual model XLSR-53-Viet, all NER models show comparable F1 scores, precision, and recall. This indicates that in addition to overall WER the models do not differ significantly in the recognition accuracy of medical NEs.
Non-cascaded models might have advantages in utilising the additional pre-training data for the downstream task. 





\section{Error Analysis}
We performed an error analysis using the best-performing models. 

\begin{figure}[h]
    \centering
    \includegraphics[width=\linewidth]{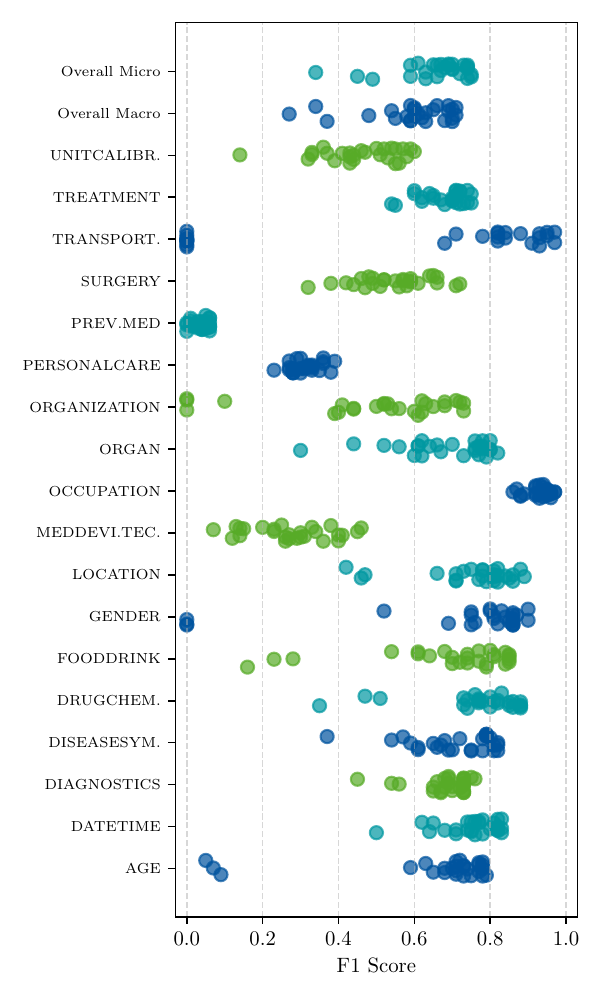}
    \caption{Scatter plot of NER results on reference text by entity types using various pre-trained language models and evaluation variants, created by Tables \ref{appx_NERresults_AGE}-\ref{appx_NERresults_OverallMicro}.}
    \label{fig:referencetext_plot}
\end{figure}

\begin{figure}[h]
    \centering
    \includegraphics[width=\linewidth]{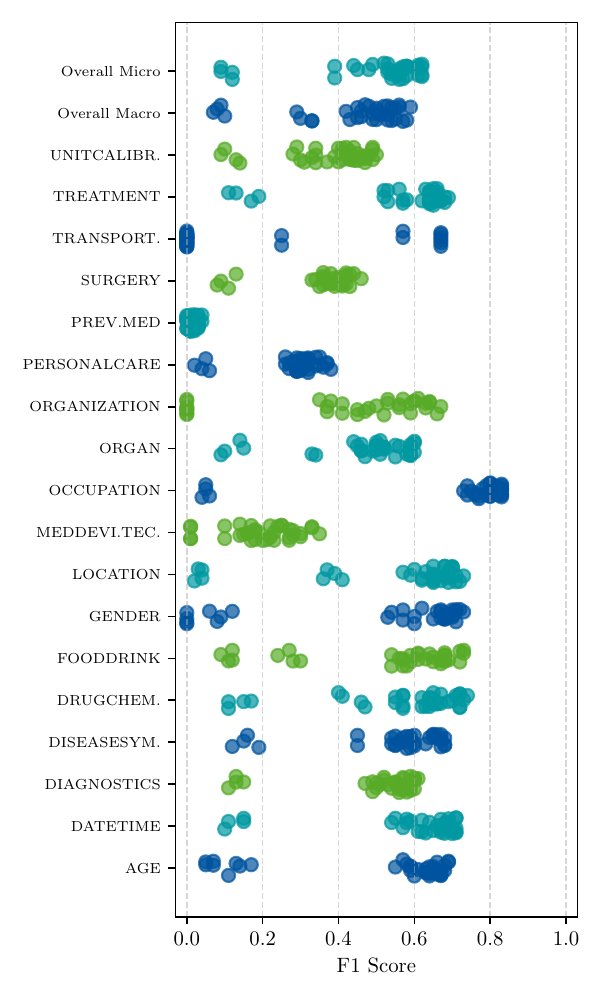}
    \caption{Scatter plot of NER results on ASR output by entity types using various pre-trained language models and evaluation variants, created by Tables \ref{appx_spk_NERresults_AGE}-\ref{appx_spk_NERresults_OverallMicro}.}
    \label{fig:asroutput_plot}
\end{figure}

\subsection{Quantitative}
We provide a detailed error analysis for each entity type across best models, utilizing three evaluation toolkits. The results are summarized in Tables \ref{appx_NERresults_AGE}-\ref{appx_NERresults_OverallMicro} and Tables \ref{appx_spk_NERresults_AGE}-\ref{appx_spk_NERresults_OverallMicro}, with corresponding visual representations in the scatter plots shown in Figure \ref{fig:referencetext_plot} and Figure \ref{fig:asroutput_plot}. 

The top NER models showed high accuracy in recognizing OCCUPATION and TRANSPORTATION entities in both reference text and ASR output. Despite TRANSPORTATION having only 35 total and 15 unique samples, the best models performed well, likely due to the semantic clarity and predictable spans of these entities.

In contrast, PREVENTIVEMED showed higher misrecognition rates, despite sufficient sample size mitigating class imbalance. This may stem from two factors. First, preventive medicine terms often overlap with general medical terminology, making it difficult for the model to distinguish them from DRUGCHEMICAL or TREATMENT concepts. For example, "vaccination" is frequently misclassified as a therapeutic intervention (TREATMENT) in sentences that describe its role in disease prevention (PREVENTIVEMED). Also, models frequently struggle to differentiate between "vaccination" and "vaccine" (DRUGCHEMICAL).
Second, preventive medicine involves long-term health strategies often expressed in non-clinical or non-standardized language, complicating entity recognition.

\subsection{Qualitative}
A common error was confusion between LOCATION and ORGANIZATION, due to the inherent ambiguity where the same entity can function as either depending on context. An organization-related entity may be labelled as LOCATION if it implies a patient visited there, but this inference requires external knowledge about the entity. Another confusion involved DRUGCHEMICAL and FOODDRINK. Both categories share similar names, descriptors, and consumption contexts (e.g., caffeine, alcohol, sugar). Insufficient context length often causes errors, especially with ambiguous terms like "vitamin," "cordyceps," or "sea daffodils," which can refer to both supplements and nutrients depending on context. Another case is DIAGNOSTICS, TREATMENT, and SURGERY. For example, a "biopsy" can be both a diagnostic and treatment, while "radiation therapy" may be linked to surgery.  

A common error in NER involves incorrect entity spans, which fall into two types: (1) correct label but wrong span, and (2) wrong label but correct span. The first type often occurs with multi-word entities in the medical domain, like ORGANIZATION, LOCATION, or DISEASESYMPTOM. For example, "high blood pressure" (B-DISEASESYMPTOM, I-DISEASESYMPTOM, I-DISEASESYMPTOM) may be misrecognized as "high blood" (B-DISEASESYMPTOM, I-DISEASESYMPTOM, O), keeping the meaning but shortening the span. The second type occurs when compound-word entities are split, such as "vagina cells" (B-ORGAN, I-ORGAN) being misrecognized as "vagina" and "cells" (B-ORGAN, B-ORGAN).

\section{Conclusion}
In this work, we present \textit{VietMed-NER} - the first spoken NER dataset in the medical domain.
Our dataset contains 18 entity types, including both conventional and newly defined entity types for real-world medical conversations.
Our results show that pre-trained multilingual models typically outperform monolingual models on both reference text and ASR output if the multilingual models are sufficiently large to learn multilingual representations. Additionally, encoders generally demonstrate better performance than seq2seq models in the NER task. Finally, while pre-trained audio data impacts ASR output, it does not significantly impact NER performance in the cascaded setting.

\section{Limitations}
\textbf{Our annotation approach}:
Our annotation approach has some advantages over the fully manual approach.
First, it allows annotators to not spend extra time tagging the entities that have been tagged in previous sentences.
Second, it prevents that annotators miss entities that have been tagged in previous sentences, improving the consistency of the entire dataset.
During our work, we experienced a faster annotation by using our approach compared to fully manual annotation.
However, in the scope of this paper, we have not done extensive experiments to give a quantitative number of how much time has been saved and the method's impact on annotation quality.

\noindent\textbf{Evaluation metrics for medical terms}: ASR system performance is commonly evaluated using WER, which quantifies the ratio of word insertion, substitution, and deletion errors in a transcript relative to the total number of spoken words. However, various spoken language understanding tasks, such as spoken NER, rely on accurately identifying key terms within transcripts. In medical ASR, it is critical to account for the disproportionate importance of medical terms in doctor-patient interactions, as they hold significantly more weight than general vocabulary, as discussed in Section \ref{sec:discussion_NEER} in the Appendix. We believe that other domain-specific spoken NER tasks follow a similar pattern. Consequently, future comprehensive investigations into evaluation metrics are needed to determine the most appropriate metric for spoken NER in the medical and other domains.


\bibliography{anthology,custom}

\clearpage 


\appendix

\onecolumn
\tableofcontents
\newpage

\twocolumn


\section{Annotation Guidelines}
\label{sec:annotation_guidelines}

This section describes annotation guidelines for annotators to follow in an attempt to have a unified and consistent gold-standard NER transcript.

\textbf{General rules}: 
\begin{itemize}
    \item If 2 or more entities overlap, label the resulting entity as the longest, including overlapping component entities.
    In other words, a full NE might contain 2 or more sub-NEs. A full NE should be tagged instead of multiple sub-NEs.
    For example:
    "bác sĩ xương khớp" (orthopedic doctor) should be tagged as a whole instead of 2 distinct NEs "bác sĩ " (doctor) and "xương khớp" (orthopedic).   
    \item We adhere to the conventional approach of annotating overlapping NE components as a whole, utilizing the BIO encoding scheme. In recent years, research on overlapping and discontinuous NER has introduced alternative annotation frameworks to solve NE overlapping, such as BILOU encoding, which represents "Beginning, Inside, and Last tokens of multi-token chunks, Unit-length chunks, and Outside" \cite{strakova2019neural, duan2017exploiting}. Another approach by \citet{li2021span} introduces a novel span-based model capable of jointly recognizing both overlapping and discontinuous entities. The model operates in two primary stages. First, entity fragments are identified by systematically traversing all possible text spans, enabling the detection of overlapping entities. Second, a relation classification step determines whether a given pair of entity fragments exhibits an overlapping or successive relationship. This approach facilitates the recognition of discontinuous entities while simultaneously verifying overlapping entities.
    \item Do not assign spaces at the beginning and at the end of entities.
    \item All words in the ASR transcript are lowercase, without punctuations and special characters.
    Treat every word as lowercase or uppercase, with or without punctuations and special characters based on the context of each utterance.
    \item Each utterance should be treated as an independent utterance.
    The additional context given by other utterances should not influence the annotation of each utterance.
\end{itemize}

\textbf{AGE}:

This entity type describes the age of a person.
\begin{itemize}
    \item Label the word "tuổi" (age) if applicable.
    For example: "tuổi trưởng thành" (mature age), "hai bảy tuổi" (twenty-seven years old).
    \item List a range of ages if applicable
    For example: "hai mươi đến ba lăm tuổi" (twenty to thirty-five years old), "dưới sáu tháng tuổi" (under six months old).
    \item Include adjectives and nouns that might describe how old a person is but don't explicitly describe gender or gender is neutral.
    For example: "chưa trưởng thành" (immature), "người già" (old person), "cụ" (sir, old).
\end{itemize}

\textbf{GENDER}:

This entity type describes the gender of a person.
\begin{itemize}
    \item Include typical entities that are widely understood to describe the gender of a person.
    For example: "nam" (male), "đàn ông" (gentleman), "phụ nữ" (woman).
    \item Include the titles and pronouns that explicitly describe a gender instead of age.
    For example: "ông" (grandfather), "bà" (grandmother), "cô" (aunt), "chú" (uncle).
\end{itemize}


\textbf{OCCUPATION}:

This entity type describes the job of a person.
\begin{itemize}
    \item Include all jobs that might be both in medical fields and non-medical fields.
    For example: "khán thính giả" (audience), "bệnh nhân" (patient), "người dân" (citizen), "chuyên gia" (expert).
    \item Include academic titles and degrees.
    For example: "thạc sĩ" (master degree holder), "tiến sĩ" (doctorate), "trưởng khoa" (dean), "chủ tịch" (president).
    \item Include a cluster of words that might describe the specializations of doctors.
    For example: "bác sĩ chuyên về rối loạn vận động" (doctor who specializes in movement disorders) instead of two distinct entities "bác sĩ" (doctor) and "rối loạn vận động" (movement disorders), "bác sĩ về parkinson" (parkinson's doctor) instead of two distinct entities "bác sĩ" (doctor) and "parkinson", "bác sĩ chuyên khoa tim mạch" (cardiovascular specialist) instead of two distinct entities "bác sĩ" (doctor) and "chuyên khoa tim mạch" (cardiovascular).
\end{itemize}

\textbf{LOCATION}:

This entity type describes a location.
\begin{itemize}
    \item Include continents, countries, regions, cities, and geographical administrative units.
    For example:
    "châu âu" (europe), "hoa kỳ" (usa), "tây tạng" (tibet), "thành phố hồ chí minh" (ho chi minh city), "tỉnh vĩnh long" (vinh long province).
    \item Label words that mean geographical administrative units if applicable.
    For example:
    "huyện" (rural district), "quận" (urban district), "đường phố" (street), "thành phố" (city).
    \item Include words that might describe public and private sites.
    For example: 
    "tại nhà" (at home), "đồng ruộng" (farm), "tiệm thuốc" (drugstore), "nhà máy" (factory), "cửa hiệu quần áo" (clothing store), "toilet" (toilet).
    \item Include words that might describe ambient environments.
    For example:
    "tại khu phố" (in the neighborhood), "tại địa phương" (in local area), "nước ngoài" (in foreign countries), "địa bàn" (area), "ngoài trời" (outside).
    \item Include words that might describe medical facilities.
    For example: "chuyên khoa tiêu hóa" (gastrointestinal room) , "icu" (intensive care unit), "trạm xá" (clinics), "phòng thí nghiệm" (laboratory).
    \item Each level of the administrative unit is a separate entity.
    \item Do not assign nationality as an entity.
    \item Locations might be misrecognized as organizations.
    Do not label places that are not clearly identified or controversial.
\end{itemize}

\textbf{DISEASESYMPTOM}:

This entity type describes a symptom or disease.
\begin{itemize}
    \item Include the complements of the disease.
    For example: "biến chứng" (side-effect), "chấn thương" (damaged), "bẩm sinh" (congenital), "di chứng" (sequelae), "bị tổn thương" (damaged), "tái phát" (relapse), "dương tính" (positive), "bệnh lý mãn tính" (chronic disease), "hội chứng" (syndrome).
    \item Include a cluster of words that might describe the severity of a disease.
    For example: "phỏng cấp độ ba" (third-degree burn), "sức đề kháng kém" (poor immune system).
    \item Mental state might also describe mental diseases or their symptoms.
    For example: "tự ti" (self-deprecation), "tình trạng lo âu" (state of anxiety), "mệt mỏi về tinh thần" (mental fatigue).
    \item Skin conditions might describe dermatosis or its symptoms.
    For example: "nám" (melasma), "da đổ dầu" (oily skin), "da khô" (dry skin), "sạm da" (dark skin).
    \item Genital conditions might describe genital diseases or their symptoms.
    For example: "có kinh" (menstruation), "có thai" (pregnant), "dậy thì sớm" (early puberty).
    \item Healthy conditions might help doctors diagnose.
    For example: "kinh nguyệt đều" (regular menstruation).
    \item Words describing physical status might also speak of symptoms or diseases.
    For example: "buồn ngủ" (sleepy), "rụng tóc" (hair loss), "còi cọc" (stunted).
    \item Words describing children's activities might also speak of pediatric symptoms or diseases. 
    For example: "quấy khóc" (fussy), "không thể giao tiếp" (unable to speak), "chậm đi" (delay walking).
    \item Medical techniques or devices might make symptoms and diseases happen.
    For example: "phẫu thuật thẩm mỹ" (cosmetic surgery).
\end{itemize}

\textbf{DRUGCHEMICAL}:

This entity type describes a bio-chemical substance or medicament.
\begin{itemize}
    \item Extraction of human or animal bodies to serve medical treatment might be referred to as a biochemical substance.
    For example: "vắcxin" (vaccine), "huyết thanh" (blood serum)
    \item Cosmetics might be referred to as chemical substances.
    For example: "kem chống nắng" (sunscreen), "kem dưỡng ẩm" (moisturizer).
    \item Food or drink serving medical treatment purposes or as a part of a chemical compound might be referred to as chemical substances.
    For example: "nấm đông trùng hạ thảo" (cordyceps), "nhân sâm" (ginseng), "nhung hươu" (deer antler).
    \item Substances extracted from cells or bodies not serving medical purposes might be referred to as bio-chemical substances.
    For example: "dịch tiêu hóa" (digestive fluids), "chất nội sinh" (endogenous substances), "mồ hôi" (sweat), "bã nhờn" (sebum).
    \item Air might be referred to as chemical substances.
    For example: "dưỡng khí" (breath air), "oxy" (oxygen).
\end{itemize}

\textbf{FOODDRINK}:

This entity type describes food and beverage.
\begin{itemize}
    \item Include food and drink that might serve nutrient purposes.
    For example: "sữa" (milk), "ngũ cốc" (cereal).
    \item Include food and drink that might be harmful to health.
    For example: "thuốc lá" (cigarette), "rượu bia" (alcohol).
    \item Include words that generally describe food and beverage.
    For example: "thực phẩm" (aliment), "thức ăn" (food).
\end{itemize}

\textbf{ORGAN}:

This entity type describes an anatomical feature, e.g. human organs, biological cells, etc.
Annotators should follow general rules.

\textbf{PERSONALCARE}:

This entity type describes a personal care procedure, e.g. hygiene routines, skin care, daily habits, etc.
\begin{itemize}
    \item Activities serving the improvement of physical, aesthetic and mental health instead of medical treatment purposes might be referred to as personal care procedures.
    For example: "ăn kiêng" (diet), "chăm sóc da" (skin care), "chăm sóc răng" (dental care).
    \item Methods serving self-improvement of speech ability in speech-language pathology might be referred to as personal care.
    For example: "tương tác ngôn ngữ" (language interaction), "huấn luyện ngôn ngữ" (language training).
\end{itemize}

\textbf{DIAGNOSTICS}:

This entity type describes a diagnostic procedure, e.g. lab tests, imaging, blood measurement, etc.
\begin{itemize}
    \item General words describing diagnostic procedures without explicitly mentioning surgery might be referred to as diagnostic produces.
    For example: "chẩn đoán" (diagnosis), "xét nghiệm" (test).
    \item Imaging methods might be referred to as diagnostic procedures instead of medical devices or techniques.
    For example: "mri" (magnetic resonance imaging), "ct" (computed tomography).
\end{itemize}

\textbf{TREATMENT}:

This entity type describes a non-surgical treatment method for diseases, e.g. physical rehabilitation, injection, psychology, etc.
\begin{itemize}
    \item Words describing methods of using bio-chemical substances as non-surgical treatment methods might be referred to as treatment methods.
    For example: "liệu pháp hoocmon" (hormone therapy), "điều trị hoocmon" (hormone treatment), "điều trị tế bào gốc" (stem cell treatment).
    \item Words describing methods of using invasive techniques as treatment methods might be referred to as treatment methods.
    For example: "hóa trị" (chemotherapy), "xạ trị" (radiotherapy).
    \item Words describing methods to improve skin conditions for treatment purposes rather than aesthetics might be referred to as treatment methods.
    For example: "phục hồi da" (skin recovery), "ức chế sự xuất sắc tố" (inhibit pigmentation).
\end{itemize}

\textbf{SURGERY}:

This entity type describes a surgical treatment method for diseases, e.g. implants, neurosurgery, invasion, etc.
\begin{itemize}
    \item Include pre-surgery procedures that might be integral parts of surgeries.
    For example: "gây mê" (anesthesia), "gây tê" (anesthetize).
    \item Include intervention procedures that might be integral parts of dental care.
    For example: "nhổ răng" (tooth extraction), "implant" (dental implant).
    \item Include intervention procedures that might be integral parts of pregnancy or genitals.
    For example: "sinh mổ" (caesarean), "cấy tránh thai" (contraceptive implant).
    \item Include intervention procedures on arteries even though they might be not integral parts of surgery.
    For example: "truyền máu" (blood transfusion), "truyền nước biển" (seawater infusion).
    \item Include neurosurgical procedures that work with brain waves even though they might be minimally invasive.
    For example: "kích thích não sâu" (dbs or deep brain stimulation).
\end{itemize}

\textbf{MEDDEVICETECHNIQUE}:

This entity type describes a medical device, instrument, bio-material and technique.

\begin{itemize}
    \item Medical devices and techniques might be confusing. 
    Annotators are strongly recommended to fully annotate CHEM., FnB, ANAT., PC, DX, TX, and SX before engaging TECH.
\end{itemize}

\textbf{UNITCALIBRATOR}:

This entity type describes a medical calibration, e.g. number of doses, calories, length, volume, etc.
\begin{itemize}
    \item Include a cluster of words that both describe the quantity and its unit. Measurements including length, distance, area, weight, heat, velocity, temperature, etc., should be explicitly tagged.
    For example: "năm milimet" (five millimeters) instead of "năm" (five) or "milimet" (millimeter).
    \item Complements to the actual quantity describing its approximation should be included.
    For example: "khoảng mười lăm phần trăm" (about fifteen percent) instead of "mười lăm phần trăm" (fifteen percent).
    \item Include words that generally describe the quantity.
    For example: "gần đủ" (close enough), "cao" (high), "rất là lớn" (very large).
    \item Include words that describe trends of quantity.
    For example: "giảm được ít nhất" (reduce at least), "mức độ gia tăng" (level increases).
\end{itemize}

\textbf{TRANSPORTATION}:

This entity type describes means of transportation or vehicles.

\textbf{DATETIME}:

This entity type describes the date and time.
\begin{itemize}
    \item Include words describing day, week, month, certain named period, season, year, etc.
    \item Include words describing a time frame.
    For example: "bây giờ" (now), "về lâu về dài" (in the long run).
    \item Include words describing the approximate time.
    For example: "nhanh nhất có thể" (as fast as possible),  "càng sớm" (as soon as possible), "từ từ" (gradually).
    \item Include words describing repetitions.
    For example: "định kỳ" (periodically).
    \item Include a cluster of words that both describe time and its complements.
    For example: "từ tháng ba trở đi" (from march onwards) instead of 3 distinct entities "từ" (from), "tháng ba" (march), and "trở đi" (onwards).
\end{itemize}

\twocolumn
\section{Discussion about Named-Entity-Error-Rate (NEER)}
\label{sec:discussion_NEER}
\subsection{Motivation of NEER}
ASR system performance is typically assessed using WER, which represents the ratio of word insertion, substitution, and deletion errors in a transcript to the total number of spoken words. 
However, various spoken language understanding tasks, such as spoken NER, depend on identifying keywords in transcripts.
Moreover, it's essential to recognize that in medical ASR, medical terms carry much higher significance in doctor-patient conversations and should not be treated equally to regular words.
KER is often used to evaluate on keywords but is not a directly comparable metric with WER.

The purpose to introduce NEER aims to bridge the gap between WER and KER.
However, it is not intended to replace WER or KER as a standard metric for evaluating domain-specific ASR performance. 
Instead, NEER serves as a complementary metric, facilitating a more in-depth analysis of ASR errors in specific domains, such as the medical field.

\subsection{Definition of WER}
WER is calculated based on the Levenshtein distance \cite{levenshtein1966levenshtein_distance}, which represents the smallest count of individual edits (insertions, deletions, or substitutions) needed to transform one word into another.

\begin{equation} \label{WER_equation}
WER = \frac{S + D + I}{N} = \frac{S + D + I}{S + D + C}
\end{equation}
where S is the number of substitutions, D is the number of deletions, I is the number of insertions, C is the number of correct words, and N is the number of words in the reference data $(N=S+D+C)$.

In other words, S is the number of replaced words.
D is the number of missed words that are not in ASR hypothesis but are in reference data.
I is the number of added words that are in ASR hypothesis but are not in reference data.
The alignment between ASR hypothesis and reference data goes from left to right.

\subsection{Definition of KER}
Like WER, KER is computed using the Levenshtein distance. 
Each ASR hypothesis is aligned with its corresponding reference data and KER is calculated based on the keyword set.

\begin{equation} \label{KER_equation}
KER = \frac{F + M}{N}
\end{equation}
where N is the number of keywords in the reference data, F is the number of falsely recognized keywords, M is the number of missed keywords.

The ASR hypothesis often exceeds the length of all keywords in the reference data, and the insertion errors caused by non-keywords may lead to a skewed result in KER.
Therefore, no insertion errors are considered while calculating KER.

\subsection{Definition of NEER}
In KER metric, N is the number of keywords in the reference data.
KER could be characterized as the average number of errors per keyword.
Nevertheless, the length of keywords may range from 1 to L (where L equals 5 in certain instances such as NER), making the average number of errors per keyword obscure.

In NEER metric, we want to evaluate on keyword-only like KER metric, while also analyzing errors per word like WER metric.
Therefore, we change N into the length of keywords (entities), which characterizes the average number of errors per word of keywords.

\subsection{Open questions on NEER}
We still leave some questions open for future work. 
First, the analysis of how each type of word error (substitutions, insertions, deletions) influences NER on top of ASR has not been conducted yet. 
Second, the empirical relationship between WER, KER, NEER, and F1 score - meaning how KER, NEER, and F1 score are affected by a varying range of WERs — has not been analyzed either.

\twocolumn
\section{Possible Applications}
\label{sec:possible_applications}
In the context of advanced in-context learning capabilities of LLMs and human-level text-to-speech technologies, the dataset, which preserves entity positional labels throughout the translation process, is applicable not only to Vietnamese but also to other languages.

To facilitate dataset translation using LLMs, we employed a concise, brief contextual instruction via GPT-3.5 Turbo, as illustrated below. 

In this example, we utilized a straightforward one-shot in-context learning approach, which successfully produced a high-quality translation while preserving all entity positional labels. We hypothesize that employing many-shot learning, extending the context length of instructions, and more advanced LLMs (e.g. GPT-4, GPT-4o) could yield even more enhanced results.

Recent research in text-to-speech (TTS) technology has achieved human-level performance in aspects such as naturalness and emotional expressiveness \cite{tan2024naturalspeech, li2024styletts, chen2023vector}. Consequently, the application of TTS to translated NER transcripts could generate realistic audio data, facilitating further research in additional languages.

\onecolumn
\begin{tcolorbox}[colback=black!5!white,colframe=black!75!black,title=gpt-3.5-turbo]

\textcolor{blue}{Instruction:}\\

Given the text: 

"thì cũng giống như \textcolor{red}{ba má} mình đã từng bị \textcolor{red}{béo phì} rồi \textcolor{red}{bị bệnh} này bệnh kia những người thân quen của mình mình biết bị"

Its coressponding named entity position is:

[ "0", "0", "0", "0", "B-GENDER", "B-GENDER", "0", "0", "0", "0", "B-DISEASESYMTOM", "I-DISEASESYMTOM", "0", "B-DISEASESYMTOM", "I-DISEASESYMTOM", "0", "0", "0", "0", "0", "0", "0", "0", "0", "0", "0", "0" ],

where "0" is none-entity. Translate to English and give the corresponding named entities.\\

\textcolor{blue}{LLM response:}\\

The given Vietnamese text translates to:

"[it is just like my mom and dad who were once obese and got this disease or that disease, people close to me, I know, have been]".

The corresponding named entities, based on the positions provided, are:

- "B-GENDER", "B-GENDER" 
\\→ \textcolor{red}{**"mom" and "dad"**} (Gender entities)

- "B-DISEASESYMTOM", "I-DISEASESYMTOM" 
\\→ \textcolor{red}{**"obese"**} (Disease/Symptom)

- "B-DISEASESYMTOM", "I-DISEASESYMTOM" 
\\→ \textcolor{red}{**"disease"**} (Disease/Symptom)

So, the translation with the named entities marked would be:

"[it is just like \textcolor{red}{**mom** and **dad**} who were once \textcolor{red}{**obese**} and got this \textcolor{red}{**disease**} or that \textcolor{red}{**disease**}, people close to me, I know, have been]".

\end{tcolorbox}

\twocolumn
\section{Details about Experimental Setups}
\label{sec:details_experimental_setups}

\subsection{Evaluation Toolkit}
We employed seqeval\footnote{https://github.com/chakki-works/seqeval} framework commonly used as a default evaluation framework by HuggingFace.
However, this framework only works for NER on reference text.
Therefore, we also employed the F1 score calculation by \citet{shon-etal-2023-slue} by using the SLUE toolkit\footnote{https://github.com/asappresearch/slue-toolkit}.
This F1 score evaluates an unordered list of NE phrase and tag pairs predicted for each sentence.
Our proposed modification of SLUE toolkit was also used and presented below.

\subsection{Modified Evaluation of SLUE toolkit}
Following pre-processing, we calculate the evaluation metrics for the ASR-NER SLUE task. 
This involves computing precision, recall, and F1-score, which provide insights into the model performance at both an individual label level (per entity) and across all labels (overall). 

We introduce a "dummy" token strategy to replace the actual NEs. 
This approach upholds the focus on the classification of entities rather than the extraction of verbatim phrases, which is suitable for cases where ASR errors might skew the recognition of entities in spoken transcripts.

Let's take an example:
\begin{itemize}
    \item Reference text: "I have a tooth pain"
    \item BIO encoding of reference text: [0, 0, 0, B-DISEASESYMTOM, I-DISEASESYMTOM]
    \item ASR output: "Has teeth pain"
    \item BIO encoding of ASR output: [0, B-DISEASESYMTOM, I-DISEASESYMTOM]
\end{itemize}

In the SLUE toolkit, the format (NE type, NE) is used to compare reference text and ASR output, e.g. (DISEASESYMTOM, "tooth pain") and (DISEASESYMTOM, "teeth pain"). 
This format gives an F1 score of 0.0 although entity type is correctly recognized.
In our "dummy" token strategy, we modify the format as (NE type, "dummy"), turning reference text and ASR output to (DISEASESYMTOM, "dummy") and (DISEASESYMTOM, "dummy") respectively. 
The modified format gives a correct F1 score of 1.0.

We compute two types of overall metrics: micro and macro averages. 
The micro average metrics aggregate the contributions of all classes to compute the average metric, while the macro average computes per-entity type metrics and averages them, without considering the frequency of each entity type. 
The micro average is therefore influenced by the class distribution and will be dominated by the performance on more frequent entity types. 
In contrast, macro averages treat all entity types equally, providing a measure of the system's performance across different types of NEs, regardless of their frequency in the dataset.

\twocolumn
\section{NER Results by Entity Types}
Tables \ref{appx_NERresults_AGE}-\ref{appx_NERresults_OverallMicro} show the results of NER on reference text by entity types using various pre-trained language models.
Tables \ref{appx_spk_NERresults_AGE}-\ref{appx_spk_NERresults_OverallMicro} show the results of NER on ASR output by entity types using various pre-trained language models and ASR models.

Figure \ref{fig:referencetext_plot} shows the scatter plot of NER results on reference text by entity types using various pre-trained language models, created by Tables \ref{appx_NERresults_AGE}-\ref{appx_NERresults_OverallMicro}.
Figure \ref{fig:asroutput_plot} shows the scatter plot of NER results on ASR output by entity types using various pre-trained language models, created by Tables \ref{appx_spk_NERresults_AGE}-\ref{appx_spk_NERresults_OverallMicro}.

\onecolumn
\begin{table}
\centering
\begin{tabular}{|c|l|l|c|c|c|}
\hline
\textbf{Entity Type} & \multicolumn{1}{c|}{\textbf{NER}} & \multicolumn{1}{c|}{\textbf{Eval. Toolkit}} & \textbf{Prec.} & \textbf{Rec.} & \textbf{F1} \\ \hline
\multirow{27}{*}{AGE} & BARTpho & Mod. SLUE & 0.70 & 0.79 & 0.74 \\ \cline{2-6} 
 & BARTpho & seqeval & 0.69 & 0.61 & 0.65 \\ \cline{2-6} 
 & BARTpho & SLUE & 0.69 & 0.78 & 0.73 \\ \cline{2-6} 
 & mBART-50 & Mod. SLUE & 0.64 & 0.81 & 0.71 \\ \cline{2-6} 
 & mBART-50 & seqeval & 0.66 & 0.53 & 0.59 \\ \cline{2-6} 
 & mBART-50 & SLUE & 0.62 & 0.79 & 0.70 \\ \cline{2-6} 
 & PhoBERT\_base & Mod. SLUE & 0.75 & 0.81 & 0.78 \\ \cline{2-6} 
 & PhoBERT\_base & seqeval & 0.76 & 0.62 & 0.68 \\ \cline{2-6} 
 & PhoBERT\_base & SLUE & 0.74 & 0.80 & 0.77 \\ \cline{2-6} 
 & PhoBERT\_base-v2 & Mod. SLUE & 0.78 & 0.78 & 0.78 \\ \cline{2-6} 
 & PhoBERT\_base-v2 & seqeval & 0.76 & 0.66 & 0.71 \\ \cline{2-6} 
 & PhoBERT\_base-v2 & SLUE & 0.77 & 0.77 & 0.77 \\ \cline{2-6} 
 & PhoBERT\_large & Mod. SLUE & 0.77 & 0.81 & 0.79 \\ \cline{2-6} 
 & PhoBERT\_large & seqeval & 0.77 & 0.66 & 0.71 \\ \cline{2-6} 
 & PhoBERT\_large & SLUE & 0.76 & 0.80 & 0.78 \\ \cline{2-6} 
 & ViDeBERTa\_base & Mod. SLUE & 0.59 & 0.05 & 0.09 \\ \cline{2-6} 
 & ViDeBERTa\_base & seqeval & 0.03 & 0.29 & 0.05 \\ \cline{2-6} 
 & ViDeBERTa\_base & SLUE & 0.50 & 0.04 & 0.07 \\ \cline{2-6} 
 & ViT5\_base & Mod. SLUE & 0.71 & 0.79 & 0.75 \\ \cline{2-6} 
 & ViT5\_base & seqeval & 0.73 & 0.63 & 0.68 \\ \cline{2-6} 
 & ViT5\_base & SLUE & 0.69 & 0.77 & 0.73 \\ \cline{2-6} 
 & XLM-R\_base & Mod. SLUE & 0.69 & 0.79 & 0.73 \\ \cline{2-6} 
 & XLM-R\_base & seqeval & 0.71 & 0.56 & 0.63 \\ \cline{2-6} 
 & XLM-R\_base & SLUE & 0.68 & 0.77 & 0.72 \\ \cline{2-6} 
 & XLM-R\_large & Mod. SLUE & 0.78 & 0.77 & 0.78 \\ \cline{2-6} 
 & XLM-R\_large & seqeval & 0.75 & 0.67 & 0.71 \\ \cline{2-6} 
 & XLM-R\_large & SLUE & 0.78 & 0.77 & 0.77 \\ \hline
\end{tabular}
\caption{NER results of \textbf{AGE} entity type (in percent) on \textbf{reference text} of test set using various pre-trained language models. Metrics shown are Precision, Recall, and overall micro/macro F1 score. Evaluation toolkits used are: seqeval, SLUE and our modified SLUE.}
\label{appx_NERresults_AGE}
\end{table}
\begin{table}
\centering
\begin{tabular}{|c|l|l|c|c|c|}
\hline
\textbf{Entity Type} & \multicolumn{1}{c|}{\textbf{NER}} & \multicolumn{1}{c|}{\textbf{Eval. Toolkit}} & \textbf{Prec.} & \textbf{Rec.} & \textbf{F1} \\ \hline
\multirow{27}{*}{DATETIME} & BARTpho & Mod. SLUE & 0.76 & 0.76 & 0.76 \\ \cline{2-6} 
 & BARTpho & seqeval & 0.62 & 0.66 & 0.64 \\ \cline{2-6} 
 & BARTpho & SLUE & 0.74 & 0.75 & 0.75 \\ \cline{2-6} 
 & mBART-50 & Mod. SLUE & 0.83 & 0.74 & 0.78 \\ \cline{2-6} 
 & mBART-50 & seqeval & 0.67 & 0.75 & 0.71 \\ \cline{2-6} 
 & mBART-50 & SLUE & 0.82 & 0.73 & 0.77 \\ \cline{2-6} 
 & PhoBERT\_base & Mod. SLUE & 0.80 & 0.83 & 0.82 \\ \cline{2-6} 
 & PhoBERT\_base & seqeval & 0.77 & 0.70 & 0.74 \\ \cline{2-6} 
 & PhoBERT\_base & SLUE & 0.80 & 0.82 & 0.81 \\ \cline{2-6} 
 & PhoBERT\_base-v2 & Mod. SLUE & 0.81 & 0.84 & 0.83 \\ \cline{2-6} 
 & PhoBERT\_base-v2 & seqeval & 0.78 & 0.73 & 0.75 \\ \cline{2-6} 
 & PhoBERT\_base-v2 & SLUE & 0.81 & 0.83 & 0.82 \\ \cline{2-6} 
 & PhoBERT\_large & Mod. SLUE & 0.83 & 0.82 & 0.83 \\ \cline{2-6} 
 & PhoBERT\_large & seqeval & 0.76 & 0.75 & 0.76 \\ \cline{2-6} 
 & PhoBERT\_large & SLUE & 0.83 & 0.81 & 0.82 \\ \cline{2-6} 
 & ViDeBERTa\_base & Mod. SLUE & 0.62 & 0.68 & 0.65 \\ \cline{2-6} 
 & ViDeBERTa\_base & seqeval & 0.58 & 0.43 & 0.50 \\ \cline{2-6} 
 & ViDeBERTa\_base & SLUE & 0.60 & 0.65 & 0.62 \\ \cline{2-6} 
 & ViT5\_base & Mod. SLUE & 0.74 & 0.77 & 0.75 \\ \cline{2-6} 
 & ViT5\_base & seqeval & 0.69 & 0.67 & 0.68 \\ \cline{2-6} 
 & ViT5\_base & SLUE & 0.72 & 0.75 & 0.74 \\ \cline{2-6} 
 & XLM-R\_base & Mod. SLUE & 0.81 & 0.78 & 0.80 \\ \cline{2-6} 
 & XLM-R\_base & seqeval & 0.72 & 0.70 & 0.71 \\ \cline{2-6} 
 & XLM-R\_base & SLUE & 0.80 & 0.77 & 0.78 \\ \cline{2-6} 
 & XLM-R\_large & Mod. SLUE & 0.85 & 0.82 & 0.83 \\ \cline{2-6} 
 & XLM-R\_large & seqeval & 0.76 & 0.77 & 0.77 \\ \cline{2-6} 
 & XLM-R\_large & SLUE & 0.84 & 0.81 & 0.82 \\ \hline
\end{tabular}
\caption{NER results of \textbf{DATETIME} entity type (in percent) on \textbf{reference text} of test set using various pre-trained language models. Metrics shown are Precision, Recall, and overall micro/macro F1 score. Evaluation toolkits used are: seqeval, SLUE and our modified SLUE.}
\label{appx_NERresults_DATETIME}
\end{table}
\begin{table}
\centering
\begin{tabular}{|c|l|l|c|c|c|}
\hline
\textbf{Entity Type} & \multicolumn{1}{c|}{\textbf{NER}} & \multicolumn{1}{c|}{\textbf{Eval. Toolkit}} & \textbf{Prec.} & \textbf{Rec.} & \textbf{F1} \\ \hline
\multirow{27}{*}{DIAGNOSTICS} & BARTpho & Mod. SLUE & 0.66 & 0.82 & 0.73 \\ \cline{2-6} 
 & BARTpho & seqeval & 0.72 & 0.65 & 0.68 \\ \cline{2-6} 
 & BARTpho & SLUE & 0.65 & 0.82 & 0.73 \\ \cline{2-6} 
 & mBART-50 & Mod. SLUE & 0.58 & 0.81 & 0.68 \\ \cline{2-6} 
 & mBART-50 & seqeval & 0.71 & 0.59 & 0.65 \\ \cline{2-6} 
 & mBART-50 & SLUE & 0.57 & 0.80 & 0.66 \\ \cline{2-6} 
 & PhoBERT\_base & Mod. SLUE & 0.66 & 0.81 & 0.73 \\ \cline{2-6} 
 & PhoBERT\_base & seqeval & 0.75 & 0.64 & 0.69 \\ \cline{2-6} 
 & PhoBERT\_base & SLUE & 0.66 & 0.80 & 0.72 \\ \cline{2-6} 
 & PhoBERT\_base-v2 & Mod. SLUE & 0.66 & 0.82 & 0.73 \\ \cline{2-6} 
 & PhoBERT\_base-v2 & seqeval & 0.77 & 0.65 & 0.70 \\ \cline{2-6} 
 & PhoBERT\_base-v2 & SLUE & 0.66 & 0.82 & 0.73 \\ \cline{2-6} 
 & PhoBERT\_large & Mod. SLUE & 0.70 & 0.83 & 0.76 \\ \cline{2-6} 
 & PhoBERT\_large & seqeval & 0.78 & 0.69 & 0.73 \\ \cline{2-6} 
 & PhoBERT\_large & SLUE & 0.69 & 0.83 & 0.75 \\ \cline{2-6} 
 & ViDeBERTa\_base & Mod. SLUE & 0.53 & 0.60 & 0.56 \\ \cline{2-6} 
 & ViDeBERTa\_base & seqeval & 0.55 & 0.39 & 0.45 \\ \cline{2-6} 
 & ViDeBERTa\_base & SLUE & 0.51 & 0.58 & 0.54 \\ \cline{2-6} 
 & ViT5\_base & Mod. SLUE & 0.60 & 0.80 & 0.69 \\ \cline{2-6} 
 & ViT5\_base & seqeval & 0.71 & 0.62 & 0.67 \\ \cline{2-6} 
 & ViT5\_base & SLUE & 0.59 & 0.78 & 0.67 \\ \cline{2-6} 
 & XLM-R\_base & Mod. SLUE & 0.61 & 0.82 & 0.70 \\ \cline{2-6} 
 & XLM-R\_base & seqeval & 0.75 & 0.58 & 0.65 \\ \cline{2-6} 
 & XLM-R\_base & SLUE & 0.60 & 0.82 & 0.69 \\ \cline{2-6} 
 & XLM-R\_large & Mod. SLUE & 0.69 & 0.78 & 0.73 \\ \cline{2-6} 
 & XLM-R\_large & seqeval & 0.78 & 0.69 & 0.73 \\ \cline{2-6} 
 & XLM-R\_large & SLUE & 0.69 & 0.78 & 0.73 \\ \hline
\end{tabular}
\caption{NER results of \textbf{DIAGNOSTICS} entity type (in percent) on \textbf{reference text} of test set using various pre-trained language models. Metrics shown are Precision, Recall, and overall micro/macro F1 score. Evaluation toolkits used are: seqeval, SLUE and our modified SLUE.}
\label{appx_NERresults_DIAGNOSTICS}
\end{table}
\begin{table}
\centering
\begin{tabular}{|c|l|l|c|c|c|}
\hline
\textbf{Entity Type} & \multicolumn{1}{c|}{\textbf{NER}} & \multicolumn{1}{c|}{\textbf{Eval. Toolkit}} & \textbf{Prec.} & \textbf{Rec.} & \textbf{F1} \\ \hline
\multirow{27}{*}{DISEASESYMTOM} & BARTpho & Mod. SLUE & 0.75 & 0.81 & 0.78 \\ \cline{2-6} 
 & BARTpho & seqeval & 0.63 & 0.60 & 0.61 \\ \cline{2-6} 
 & BARTpho & SLUE & 0.73 & 0.78 & 0.75 \\ \cline{2-6} 
 & mBART-50 & Mod. SLUE & 0.71 & 0.73 & 0.72 \\ \cline{2-6} 
 & mBART-50 & seqeval & 0.57 & 0.56 & 0.57 \\ \cline{2-6} 
 & mBART-50 & SLUE & 0.68 & 0.69 & 0.69 \\ \cline{2-6} 
 & PhoBERT\_base & Mod. SLUE & 0.79 & 0.83 & 0.81 \\ \cline{2-6} 
 & PhoBERT\_base & seqeval & 0.70 & 0.60 & 0.65 \\ \cline{2-6} 
 & PhoBERT\_base & SLUE & 0.77 & 0.80 & 0.78 \\ \cline{2-6} 
 & PhoBERT\_base-v2 & Mod. SLUE & 0.79 & 0.85 & 0.82 \\ \cline{2-6} 
 & PhoBERT\_base-v2 & seqeval & 0.73 & 0.61 & 0.66 \\ \cline{2-6} 
 & PhoBERT\_base-v2 & SLUE & 0.77 & 0.82 & 0.79 \\ \cline{2-6} 
 & PhoBERT\_large & Mod. SLUE & 0.82 & 0.81 & 0.82 \\ \cline{2-6} 
 & PhoBERT\_large & seqeval & 0.71 & 0.64 & 0.67 \\ \cline{2-6} 
 & PhoBERT\_large & SLUE & 0.79 & 0.79 & 0.79 \\ \cline{2-6} 
 & ViDeBERTa\_base & Mod. SLUE & 0.68 & 0.52 & 0.59 \\ \cline{2-6} 
 & ViDeBERTa\_base & seqeval & 0.35 & 0.38 & 0.37 \\ \cline{2-6} 
 & ViDeBERTa\_base & SLUE & 0.62 & 0.47 & 0.54 \\ \cline{2-6} 
 & ViT5\_base & Mod. SLUE & 0.78 & 0.84 & 0.81 \\ \cline{2-6} 
 & ViT5\_base & seqeval & 0.73 & 0.67 & 0.70 \\ \cline{2-6} 
 & ViT5\_base & SLUE & 0.77 & 0.82 & 0.79 \\ \cline{2-6} 
 & XLM-R\_base & Mod. SLUE & 0.75 & 0.83 & 0.79 \\ \cline{2-6} 
 & XLM-R\_base & seqeval & 0.69 & 0.56 & 0.61 \\ \cline{2-6} 
 & XLM-R\_base & SLUE & 0.72 & 0.79 & 0.75 \\ \cline{2-6} 
 & XLM-R\_large & Mod. SLUE & 0.83 & 0.81 & 0.82 \\ \cline{2-6} 
 & XLM-R\_large & seqeval & 0.70 & 0.66 & 0.68 \\ \cline{2-6} 
 & XLM-R\_large & SLUE & 0.81 & 0.79 & 0.80 \\ \hline
\end{tabular}
\caption{NER results of \textbf{DISEASESYMTOM} entity type (in percent) on \textbf{reference text} of test set using various pre-trained language models. Metrics shown are Precision, Recall, and overall micro/macro F1 score. Evaluation toolkits used are: seqeval, SLUE and our modified SLUE.}
\label{appx_NERresults_DISEASESYMTOM}
\end{table}
\begin{table}
\centering
\begin{tabular}{|c|l|l|c|c|c|}
\hline
\textbf{Entity Type} & \multicolumn{1}{c|}{\textbf{NER}} & \multicolumn{1}{c|}{\textbf{Eval. Toolkit}} & \textbf{Prec.} & \textbf{Rec.} & \textbf{F1} \\ \hline
\multirow{27}{*}{DRUGCHEMICAL} & BARTpho & Mod. SLUE & 0.73 & 0.82 & 0.77 \\ \cline{2-6} 
 & BARTpho & seqeval & 0.75 & 0.70 & 0.73 \\ \cline{2-6} 
 & BARTpho & SLUE & 0.73 & 0.81 & 0.77 \\ \cline{2-6} 
 & mBART-50 & Mod. SLUE & 0.79 & 0.69 & 0.74 \\ \cline{2-6} 
 & mBART-50 & seqeval & 0.71 & 0.76 & 0.73 \\ \cline{2-6} 
 & mBART-50 & SLUE & 0.79 & 0.69 & 0.74 \\ \cline{2-6} 
 & PhoBERT\_base & Mod. SLUE & 0.80 & 0.93 & 0.86 \\ \cline{2-6} 
 & PhoBERT\_base & seqeval & 0.91 & 0.77 & 0.83 \\ \cline{2-6} 
 & PhoBERT\_base & SLUE & 0.79 & 0.93 & 0.86 \\ \cline{2-6} 
 & PhoBERT\_base-v2 & Mod. SLUE & 0.83 & 0.93 & 0.88 \\ \cline{2-6} 
 & PhoBERT\_base-v2 & seqeval & 0.91 & 0.80 & 0.85 \\ \cline{2-6} 
 & PhoBERT\_base-v2 & SLUE & 0.83 & 0.93 & 0.88 \\ \cline{2-6} 
 & PhoBERT\_large & Mod. SLUE & 0.74 & 0.93 & 0.82 \\ \cline{2-6} 
 & PhoBERT\_large & seqeval & 0.93 & 0.72 & 0.81 \\ \cline{2-6} 
 & PhoBERT\_large & SLUE & 0.74 & 0.93 & 0.82 \\ \cline{2-6} 
 & ViDeBERTa\_base & Mod. SLUE & 0.65 & 0.41 & 0.51 \\ \cline{2-6} 
 & ViDeBERTa\_base & seqeval & 0.31 & 0.41 & 0.35 \\ \cline{2-6} 
 & ViDeBERTa\_base & SLUE & 0.60 & 0.38 & 0.47 \\ \cline{2-6} 
 & ViT5\_base & Mod. SLUE & 0.75 & 0.87 & 0.80 \\ \cline{2-6} 
 & ViT5\_base & seqeval & 0.83 & 0.74 & 0.78 \\ \cline{2-6} 
 & ViT5\_base & SLUE & 0.75 & 0.86 & 0.80 \\ \cline{2-6} 
 & XLM-R\_base & Mod. SLUE & 0.81 & 0.74 & 0.77 \\ \cline{2-6} 
 & XLM-R\_base & seqeval & 0.76 & 0.76 & 0.76 \\ \cline{2-6} 
 & XLM-R\_base & SLUE & 0.80 & 0.74 & 0.77 \\ \cline{2-6} 
 & XLM-R\_large & Mod. SLUE & 0.85 & 0.92 & 0.88 \\ \cline{2-6} 
 & XLM-R\_large & seqeval & 0.91 & 0.79 & 0.85 \\ \cline{2-6} 
 & XLM-R\_large & SLUE & 0.85 & 0.92 & 0.88 \\ \hline
\end{tabular}
\caption{NER results of \textbf{DRUGCHEMICAL} entity type (in percent) on \textbf{reference text} of test set using various pre-trained language models. Metrics shown are Precision, Recall, and overall micro/macro F1 score. Evaluation toolkits used are: seqeval, SLUE and our modified SLUE.}
\label{appx_NERresults_DRUGCHEMICAL}
\end{table}
\begin{table}
\centering
\begin{tabular}{|c|l|l|c|c|c|}
\hline
\textbf{Entity Type} & \multicolumn{1}{c|}{\textbf{NER}} & \multicolumn{1}{c|}{\textbf{Eval. Toolkit}} & \textbf{Prec.} & \textbf{Rec.} & \textbf{F1} \\ \hline
\multirow{27}{*}{FOODDRINK} & BARTpho & Mod. SLUE & 0.67 & 0.83 & 0.74 \\ \cline{2-6} 
 & BARTpho & seqeval & 0.69 & 0.59 & 0.64 \\ \cline{2-6} 
 & BARTpho & SLUE & 0.67 & 0.83 & 0.74 \\ \cline{2-6} 
 & mBART-50 & Mod. SLUE & 0.56 & 0.67 & 0.61 \\ \cline{2-6} 
 & mBART-50 & seqeval & 0.57 & 0.52 & 0.54 \\ \cline{2-6} 
 & mBART-50 & SLUE & 0.56 & 0.67 & 0.61 \\ \cline{2-6} 
 & PhoBERT\_base & Mod. SLUE & 0.78 & 0.84 & 0.81 \\ \cline{2-6} 
 & PhoBERT\_base & seqeval & 0.74 & 0.67 & 0.70 \\ \cline{2-6} 
 & PhoBERT\_base & SLUE & 0.78 & 0.84 & 0.81 \\ \cline{2-6} 
 & PhoBERT\_base-v2 & Mod. SLUE & 0.82 & 0.89 & 0.85 \\ \cline{2-6} 
 & PhoBERT\_base-v2 & seqeval & 0.83 & 0.76 & 0.79 \\ \cline{2-6} 
 & PhoBERT\_base-v2 & SLUE & 0.82 & 0.89 & 0.85 \\ \cline{2-6} 
 & PhoBERT\_large & Mod. SLUE & 0.80 & 0.91 & 0.85 \\ \cline{2-6} 
 & PhoBERT\_large & seqeval & 0.85 & 0.75 & 0.80 \\ \cline{2-6} 
 & PhoBERT\_large & SLUE & 0.80 & 0.91 & 0.85 \\ \cline{2-6} 
 & ViDeBERTa\_base & Mod. SLUE & 0.22 & 0.38 & 0.28 \\ \cline{2-6} 
 & ViDeBERTa\_base & seqeval & 0.24 & 0.12 & 0.16 \\ \cline{2-6} 
 & ViDeBERTa\_base & SLUE & 0.19 & 0.32 & 0.23 \\ \cline{2-6} 
 & ViT5\_base & Mod. SLUE & 0.70 & 0.86 & 0.77 \\ \cline{2-6} 
 & ViT5\_base & seqeval & 0.75 & 0.66 & 0.70 \\ \cline{2-6} 
 & ViT5\_base & SLUE & 0.70 & 0.86 & 0.77 \\ \cline{2-6} 
 & XLM-R\_base & Mod. SLUE & 0.64 & 0.88 & 0.74 \\ \cline{2-6} 
 & XLM-R\_base & seqeval & 0.82 & 0.58 & 0.68 \\ \cline{2-6} 
 & XLM-R\_base & SLUE & 0.62 & 0.85 & 0.72 \\ \cline{2-6} 
 & XLM-R\_large & Mod. SLUE & 0.88 & 0.81 & 0.84 \\ \cline{2-6} 
 & XLM-R\_large & seqeval & 0.75 & 0.83 & 0.79 \\ \cline{2-6} 
 & XLM-R\_large & SLUE & 0.88 & 0.81 & 0.84 \\ \hline
\end{tabular}
\caption{NER results of \textbf{FOODDRINK} entity type (in percent) on \textbf{reference text} of test set using various pre-trained language models. Metrics shown are Precision, Recall, and overall micro/macro F1 score. Evaluation toolkits used are: seqeval, SLUE and our modified SLUE.}
\label{appx_NERresults_FOODDRINK}
\end{table}
\begin{table}
\centering
\begin{tabular}{|c|l|l|c|c|c|}
\hline
\textbf{Entity Type} & \multicolumn{1}{c|}{\textbf{NER}} & \multicolumn{1}{c|}{\textbf{Eval. Toolkit}} & \textbf{Prec.} & \textbf{Rec.} & \textbf{F1} \\ \hline
\multirow{27}{*}{GENDER} & BARTpho & Mod. SLUE & 0.83 & 0.90 & 0.86 \\ \cline{2-6} 
 & BARTpho & seqeval & 0.85 & 0.78 & 0.81 \\ \cline{2-6} 
 & BARTpho & SLUE & 0.82 & 0.90 & 0.86 \\ \cline{2-6} 
 & mBART-50 & Mod. SLUE & 0.83 & 0.87 & 0.85 \\ \cline{2-6} 
 & mBART-50 & seqeval & 0.77 & 0.75 & 0.76 \\ \cline{2-6} 
 & mBART-50 & SLUE & 0.82 & 0.86 & 0.84 \\ \cline{2-6} 
 & PhoBERT\_base & Mod. SLUE & 0.83 & 0.90 & 0.86 \\ \cline{2-6} 
 & PhoBERT\_base & seqeval & 0.76 & 0.74 & 0.75 \\ \cline{2-6} 
 & PhoBERT\_base & SLUE & 0.83 & 0.90 & 0.86 \\ \cline{2-6} 
 & PhoBERT\_base-v2 & Mod. SLUE & 0.84 & 0.89 & 0.86 \\ \cline{2-6} 
 & PhoBERT\_base-v2 & seqeval & 0.83 & 0.77 & 0.80 \\ \cline{2-6} 
 & PhoBERT\_base-v2 & SLUE & 0.84 & 0.89 & 0.86 \\ \cline{2-6} 
 & PhoBERT\_large & Mod. SLUE & 0.83 & 0.90 & 0.87 \\ \cline{2-6} 
 & PhoBERT\_large & seqeval & 0.87 & 0.79 & 0.83 \\ \cline{2-6} 
 & PhoBERT\_large & SLUE & 0.83 & 0.90 & 0.86 \\ \cline{2-6} 
 & ViDeBERTa\_base & Mod. SLUE & 0.00 & 0.00 & 0.00 \\ \cline{2-6} 
 & ViDeBERTa\_base & seqeval & 0.00 & 0.00 & 0.00 \\ \cline{2-6} 
 & ViDeBERTa\_base & SLUE & 0.00 & 0.00 & 0.00 \\ \cline{2-6} 
 & ViT5\_base & Mod. SLUE & 0.81 & 0.82 & 0.82 \\ \cline{2-6} 
 & ViT5\_base & seqeval & 0.67 & 0.71 & 0.69 \\ \cline{2-6} 
 & ViT5\_base & SLUE & 0.81 & 0.82 & 0.81 \\ \cline{2-6} 
 & XLM-R\_base & Mod. SLUE & 0.79 & 0.72 & 0.75 \\ \cline{2-6} 
 & XLM-R\_base & seqeval & 0.56 & 0.49 & 0.52 \\ \cline{2-6} 
 & XLM-R\_base & SLUE & 0.78 & 0.71 & 0.75 \\ \cline{2-6} 
 & XLM-R\_large & Mod. SLUE & 0.91 & 0.90 & 0.90 \\ \cline{2-6} 
 & XLM-R\_large & seqeval & 0.82 & 0.79 & 0.80 \\ \cline{2-6} 
 & XLM-R\_large & SLUE & 0.91 & 0.89 & 0.90 \\ \hline
\end{tabular}
\caption{NER results of \textbf{GENDER} entity type (in percent) on \textbf{reference text} of test set using various pre-trained language models. Metrics shown are Precision, Recall, and overall micro/macro F1 score. Evaluation toolkits used are: seqeval, SLUE and our modified SLUE.}
\label{appx_NERresults_GENDER}
\end{table}
\begin{table}
\centering
\begin{tabular}{|c|l|l|c|c|c|}
\hline
\textbf{Entity Type} & \multicolumn{1}{c|}{\textbf{NER}} & \multicolumn{1}{c|}{\textbf{Eval. Toolkit}} & \textbf{Prec.} & \textbf{Rec.} & \textbf{F1} \\ \hline
\multirow{27}{*}{LOCATION} & BARTpho & Mod. SLUE & 0.77 & 0.88 & 0.82 \\ \cline{2-6} 
 & BARTpho & seqeval & 0.75 & 0.68 & 0.71 \\ \cline{2-6} 
 & BARTpho & SLUE & 0.76 & 0.87 & 0.81 \\ \cline{2-6} 
 & mBART-50 & Mod. SLUE & 0.74 & 0.83 & 0.78 \\ \cline{2-6} 
 & mBART-50 & seqeval & 0.73 & 0.68 & 0.71 \\ \cline{2-6} 
 & mBART-50 & SLUE & 0.74 & 0.82 & 0.78 \\ \cline{2-6} 
 & PhoBERT\_base & Mod. SLUE & 0.75 & 0.92 & 0.82 \\ \cline{2-6} 
 & PhoBERT\_base & seqeval & 0.81 & 0.63 & 0.71 \\ \cline{2-6} 
 & PhoBERT\_base & SLUE & 0.74 & 0.91 & 0.81 \\ \cline{2-6} 
 & PhoBERT\_base-v2 & Mod. SLUE & 0.78 & 0.95 & 0.86 \\ \cline{2-6} 
 & PhoBERT\_base-v2 & seqeval & 0.86 & 0.69 & 0.77 \\ \cline{2-6} 
 & PhoBERT\_base-v2 & SLUE & 0.78 & 0.94 & 0.86 \\ \cline{2-6} 
 & PhoBERT\_large & Mod. SLUE & 0.80 & 0.91 & 0.85 \\ \cline{2-6} 
 & PhoBERT\_large & seqeval & 0.82 & 0.69 & 0.75 \\ \cline{2-6} 
 & PhoBERT\_large & SLUE & 0.80 & 0.90 & 0.84 \\ \cline{2-6} 
 & ViDeBERTa\_base & Mod. SLUE & 0.78 & 0.33 & 0.47 \\ \cline{2-6} 
 & ViDeBERTa\_base & seqeval & 0.32 & 0.60 & 0.42 \\ \cline{2-6} 
 & ViDeBERTa\_base & SLUE & 0.77 & 0.33 & 0.46 \\ \cline{2-6} 
 & ViT5\_base & Mod. SLUE & 0.74 & 0.92 & 0.82 \\ \cline{2-6} 
 & ViT5\_base & seqeval & 0.83 & 0.65 & 0.73 \\ \cline{2-6} 
 & ViT5\_base & SLUE & 0.73 & 0.92 & 0.81 \\ \cline{2-6} 
 & XLM-R\_base & Mod. SLUE & 0.75 & 0.84 & 0.79 \\ \cline{2-6} 
 & XLM-R\_base & seqeval & 0.75 & 0.59 & 0.66 \\ \cline{2-6} 
 & XLM-R\_base & SLUE & 0.74 & 0.82 & 0.78 \\ \cline{2-6} 
 & XLM-R\_large & Mod. SLUE & 0.85 & 0.93 & 0.89 \\ \cline{2-6} 
 & XLM-R\_large & seqeval & 0.87 & 0.75 & 0.81 \\ \cline{2-6} 
 & XLM-R\_large & SLUE & 0.85 & 0.93 & 0.88 \\ \hline
\end{tabular}
\caption{NER results of \textbf{LOCATION} entity type (in percent) on \textbf{reference text} of test set using various pre-trained language models. Metrics shown are Precision, Recall, and overall micro/macro F1 score. Evaluation toolkits used are: seqeval, SLUE and our modified SLUE.}
\label{appx_NERresults_LOCATION}
\end{table}
\begin{table}
\centering
\begin{tabular}{|c|l|l|c|c|c|}
\hline
\textbf{Entity Type} & \multicolumn{1}{c|}{\textbf{NER}} & \multicolumn{1}{c|}{\textbf{Eval. Toolkit}} & \textbf{Prec.} & \textbf{Rec.} & \textbf{F1} \\ \hline
\multirow{27}{*}{LOCATION} & BARTpho & Mod. SLUE & 0.54 & 0.21 & 0.30 \\ \cline{2-6} 
 & BARTpho & seqeval & 0.16 & 0.39 & 0.23 \\ \cline{2-6} 
 & BARTpho & SLUE & 0.48 & 0.18 & 0.26 \\ \cline{2-6} 
 & mBART-50 & Mod. SLUE & 0.45 & 0.09 & 0.15 \\ \cline{2-6} 
 & mBART-50 & seqeval & 0.08 & 0.35 & 0.13 \\ \cline{2-6} 
 & mBART-50 & SLUE & 0.36 & 0.07 & 0.12 \\ \cline{2-6} 
 & PhoBERT\_base & Mod. SLUE & 0.55 & 0.38 & 0.45 \\ \cline{2-6} 
 & PhoBERT\_base & seqeval & 0.24 & 0.27 & 0.26 \\ \cline{2-6} 
 & PhoBERT\_base & SLUE & 0.49 & 0.34 & 0.40 \\ \cline{2-6} 
 & PhoBERT\_base-v2 & Mod. SLUE & 0.59 & 0.38 & 0.46 \\ \cline{2-6} 
 & PhoBERT\_base-v2 & seqeval & 0.27 & 0.35 & 0.31 \\ \cline{2-6} 
 & PhoBERT\_base-v2 & SLUE & 0.53 & 0.34 & 0.41 \\ \cline{2-6} 
 & PhoBERT\_large & Mod. SLUE & 0.61 & 0.30 & 0.40 \\ \cline{2-6} 
 & PhoBERT\_large & seqeval & 0.22 & 0.35 & 0.27 \\ \cline{2-6} 
 & PhoBERT\_large & SLUE & 0.55 & 0.27 & 0.36 \\ \cline{2-6} 
 & ViDeBERTa\_base & Mod. SLUE & 0.34 & 0.14 & 0.20 \\ \cline{2-6} 
 & ViDeBERTa\_base & seqeval & 0.06 & 0.09 & 0.07 \\ \cline{2-6} 
 & ViDeBERTa\_base & SLUE & 0.24 & 0.10 & 0.14 \\ \cline{2-6} 
 & ViT5\_base & Mod. SLUE & 0.52 & 0.21 & 0.30 \\ \cline{2-6} 
 & ViT5\_base & seqeval & 0.16 & 0.37 & 0.23 \\ \cline{2-6} 
 & ViT5\_base & SLUE & 0.47 & 0.19 & 0.27 \\ \cline{2-6} 
 & XLM-R\_base & Mod. SLUE & 0.48 & 0.27 & 0.34 \\ \cline{2-6} 
 & XLM-R\_base & seqeval & 0.12 & 0.18 & 0.14 \\ \cline{2-6} 
 & XLM-R\_base & SLUE & 0.41 & 0.23 & 0.29 \\ \cline{2-6} 
 & XLM-R\_large & Mod. SLUE & 0.58 & 0.28 & 0.38 \\ \cline{2-6} 
 & XLM-R\_large & seqeval & 0.20 & 0.32 & 0.25 \\ \cline{2-6} 
 & XLM-R\_large & SLUE & 0.51 & 0.25 & 0.33 \\ \hline
\end{tabular}
\caption{NER results of \textbf{MEDDEVICETECHNIQUE} entity type (in percent) on \textbf{reference text} of test set using various pre-trained language models. Metrics shown are Precision, Recall, and overall micro/macro F1 score. Evaluation toolkits used are: seqeval, SLUE and our modified SLUE.}
\label{appx_NERresults_MEDDEVICETECHNIQUE}
\end{table}
\begin{table}
\centering
\begin{tabular}{|c|l|l|c|c|c|}
\hline
\textbf{Entity Type} & \multicolumn{1}{c|}{\textbf{NER}} & \multicolumn{1}{c|}{\textbf{Eval. Toolkit}} & \textbf{Prec.} & \textbf{Rec.} & \textbf{F1} \\ \hline
\multirow{27}{*}{OCCUPATION} & BARTpho & Mod. SLUE & 0.91 & 0.93 & 0.92 \\ \cline{2-6} 
 & BARTpho & seqeval & 0.87 & 0.87 & 0.87 \\ \cline{2-6} 
 & BARTpho & SLUE & 0.91 & 0.92 & 0.92 \\ \cline{2-6} 
 & mBART-50 & Mod. SLUE & 0.97 & 0.93 & 0.95 \\ \cline{2-6} 
 & mBART-50 & seqeval & 0.91 & 0.95 & 0.93 \\ \cline{2-6} 
 & mBART-50 & SLUE & 0.97 & 0.93 & 0.95 \\ \cline{2-6} 
 & PhoBERT\_base & Mod. SLUE & 0.95 & 0.96 & 0.95 \\ \cline{2-6} 
 & PhoBERT\_base & seqeval & 0.94 & 0.92 & 0.93 \\ \cline{2-6} 
 & PhoBERT\_base & SLUE & 0.95 & 0.96 & 0.95 \\ \cline{2-6} 
 & PhoBERT\_base-v2 & Mod. SLUE & 0.96 & 0.96 & 0.96 \\ \cline{2-6} 
 & PhoBERT\_base-v2 & seqeval & 0.95 & 0.93 & 0.94 \\ \cline{2-6} 
 & PhoBERT\_base-v2 & SLUE & 0.96 & 0.96 & 0.96 \\ \cline{2-6} 
 & PhoBERT\_large & Mod. SLUE & 0.97 & 0.95 & 0.96 \\ \cline{2-6} 
 & PhoBERT\_large & seqeval & 0.93 & 0.94 & 0.94 \\ \cline{2-6} 
 & PhoBERT\_large & SLUE & 0.97 & 0.95 & 0.96 \\ \cline{2-6} 
 & ViDeBERTa\_base & Mod. SLUE & 0.96 & 0.82 & 0.89 \\ \cline{2-6} 
 & ViDeBERTa\_base & seqeval & 0.81 & 0.91 & 0.86 \\ \cline{2-6} 
 & ViDeBERTa\_base & SLUE & 0.96 & 0.81 & 0.88 \\ \cline{2-6} 
 & ViT5\_base & Mod. SLUE & 0.95 & 0.93 & 0.94 \\ \cline{2-6} 
 & ViT5\_base & seqeval & 0.92 & 0.94 & 0.93 \\ \cline{2-6} 
 & ViT5\_base & SLUE & 0.95 & 0.93 & 0.94 \\ \cline{2-6} 
 & XLM-R\_base & Mod. SLUE & 0.88 & 0.96 & 0.92 \\ \cline{2-6} 
 & XLM-R\_base & seqeval & 0.93 & 0.83 & 0.88 \\ \cline{2-6} 
 & XLM-R\_base & SLUE & 0.88 & 0.96 & 0.92 \\ \cline{2-6} 
 & XLM-R\_large & Mod. SLUE & 0.97 & 0.97 & 0.97 \\ \cline{2-6} 
 & XLM-R\_large & seqeval & 0.95 & 0.95 & 0.95 \\ \cline{2-6} 
 & XLM-R\_large & SLUE & 0.97 & 0.97 & 0.97 \\ \hline
\end{tabular}
\caption{NER results of \textbf{OCCUPATION} entity type (in percent) on \textbf{reference text} of test set using various pre-trained language models. Metrics shown are Precision, Recall, and overall micro/macro F1 score. Evaluation toolkits used are: seqeval, SLUE and our modified SLUE.}
\label{appx_NERresults_OCCUPATION}
\end{table}
\begin{table}
\centering
\begin{tabular}{|c|l|l|c|c|c|}
\hline
\textbf{Entity Type} & \multicolumn{1}{c|}{\textbf{NER}} & \multicolumn{1}{c|}{\textbf{Eval. Toolkit}} & \textbf{Prec.} & \textbf{Rec.} & \textbf{F1} \\ \hline
\multirow{27}{*}{ORGAN} & BARTpho & Mod. SLUE & 0.72 & 0.80 & 0.76 \\ \cline{2-6} 
 & BARTpho & seqeval & 0.63 & 0.56 & 0.60 \\ \cline{2-6} 
 & BARTpho & SLUE & 0.70 & 0.77 & 0.73 \\ \cline{2-6} 
 & mBART-50 & Mod. SLUE & 0.67 & 0.74 & 0.70 \\ \cline{2-6} 
 & mBART-50 & seqeval & 0.61 & 0.51 & 0.56 \\ \cline{2-6} 
 & mBART-50 & SLUE & 0.64 & 0.71 & 0.67 \\ \cline{2-6} 
 & PhoBERT\_base & Mod. SLUE & 0.74 & 0.86 & 0.79 \\ \cline{2-6} 
 & PhoBERT\_base & seqeval & 0.69 & 0.55 & 0.61 \\ \cline{2-6} 
 & PhoBERT\_base & SLUE & 0.71 & 0.83 & 0.77 \\ \cline{2-6} 
 & PhoBERT\_base-v2 & Mod. SLUE & 0.74 & 0.88 & 0.80 \\ \cline{2-6} 
 & PhoBERT\_base-v2 & seqeval & 0.70 & 0.55 & 0.62 \\ \cline{2-6} 
 & PhoBERT\_base-v2 & SLUE & 0.72 & 0.86 & 0.78 \\ \cline{2-6} 
 & PhoBERT\_large & Mod. SLUE & 0.73 & 0.87 & 0.80 \\ \cline{2-6} 
 & PhoBERT\_large & seqeval & 0.71 & 0.55 & 0.62 \\ \cline{2-6} 
 & PhoBERT\_large & SLUE & 0.71 & 0.85 & 0.77 \\ \cline{2-6} 
 & ViDeBERTa\_base & Mod. SLUE & 0.53 & 0.51 & 0.52 \\ \cline{2-6} 
 & ViDeBERTa\_base & seqeval & 0.32 & 0.27 & 0.30 \\ \cline{2-6} 
 & ViDeBERTa\_base & SLUE & 0.45 & 0.44 & 0.44 \\ \cline{2-6} 
 & ViT5\_base & Mod. SLUE & 0.71 & 0.85 & 0.78 \\ \cline{2-6} 
 & ViT5\_base & seqeval & 0.70 & 0.58 & 0.64 \\ \cline{2-6} 
 & ViT5\_base & SLUE & 0.70 & 0.84 & 0.76 \\ \cline{2-6} 
 & XLM-R\_base & Mod. SLUE & 0.71 & 0.87 & 0.78 \\ \cline{2-6} 
 & XLM-R\_base & seqeval & 0.70 & 0.55 & 0.61 \\ \cline{2-6} 
 & XLM-R\_base & SLUE & 0.69 & 0.85 & 0.76 \\ \cline{2-6} 
 & XLM-R\_large & Mod. SLUE & 0.77 & 0.87 & 0.82 \\ \cline{2-6} 
 & XLM-R\_large & seqeval & 0.75 & 0.60 & 0.66 \\ \cline{2-6} 
 & XLM-R\_large & SLUE & 0.75 & 0.85 & 0.80 \\ \hline
\end{tabular}
\caption{NER results of \textbf{ORGAN} entity type (in percent) on \textbf{reference text} of test set using various pre-trained language models. Metrics shown are Precision, Recall, and overall micro/macro F1 score. Evaluation toolkits used are: seqeval, SLUE and our modified SLUE.}
\label{appx_NERresults_ORGAN}
\end{table}
\begin{table}
\centering
\begin{tabular}{|c|l|l|c|c|c|}
\hline
\textbf{Entity Type} & \multicolumn{1}{c|}{\textbf{NER}} & \multicolumn{1}{c|}{\textbf{Eval. Toolkit}} & \textbf{Prec.} & \textbf{Rec.} & \textbf{F1} \\ \hline
\multirow{27}{*}{ORGANIZATION} & BARTpho & Mod. SLUE & 0.73 & 0.54 & 0.62 \\ \cline{2-6} 
 & BARTpho & seqeval & 0.47 & 0.68 & 0.56 \\ \cline{2-6} 
 & BARTpho & SLUE & 0.71 & 0.53 & 0.61 \\ \cline{2-6} 
 & mBART-50 & Mod. SLUE & 0.77 & 0.31 & 0.44 \\ \cline{2-6} 
 & mBART-50 & seqeval & 0.28 & 0.70 & 0.40 \\ \cline{2-6} 
 & mBART-50 & SLUE & 0.77 & 0.31 & 0.44 \\ \cline{2-6} 
 & PhoBERT\_base & Mod. SLUE & 0.71 & 0.57 & 0.63 \\ \cline{2-6} 
 & PhoBERT\_base & seqeval & 0.51 & 0.56 & 0.53 \\ \cline{2-6} 
 & PhoBERT\_base & SLUE & 0.70 & 0.56 & 0.62 \\ \cline{2-6} 
 & PhoBERT\_base-v2 & Mod. SLUE & 0.76 & 0.71 & 0.73 \\ \cline{2-6} 
 & PhoBERT\_base-v2 & seqeval & 0.59 & 0.60 & 0.60 \\ \cline{2-6} 
 & PhoBERT\_base-v2 & SLUE & 0.74 & 0.70 & 0.72 \\ \cline{2-6} 
 & PhoBERT\_large & Mod. SLUE & 0.71 & 0.65 & 0.68 \\ \cline{2-6} 
 & PhoBERT\_large & seqeval & 0.59 & 0.50 & 0.54 \\ \cline{2-6} 
 & PhoBERT\_large & SLUE & 0.71 & 0.65 & 0.68 \\ \cline{2-6} 
 & ViDeBERTa\_base & Mod. SLUE & 0.00 & 0.00 & 0.00 \\ \cline{2-6} 
 & ViDeBERTa\_base & seqeval & 0.00 & 0.00 & 0.00 \\ \cline{2-6} 
 & ViDeBERTa\_base & SLUE & 0.00 & 0.00 & 0.00 \\ \cline{2-6} 
 & ViT5\_base & Mod. SLUE & 0.94 & 0.36 & 0.52 \\ \cline{2-6} 
 & ViT5\_base & seqeval & 0.34 & 0.91 & 0.50 \\ \cline{2-6} 
 & ViT5\_base & SLUE & 0.94 & 0.36 & 0.52 \\ \cline{2-6} 
 & XLM-R\_base & Mod. SLUE & 0.59 & 0.31 & 0.41 \\ \cline{2-6} 
 & XLM-R\_base & seqeval & 0.08 & 0.12 & 0.10 \\ \cline{2-6} 
 & XLM-R\_base & SLUE & 0.56 & 0.29 & 0.39 \\ \cline{2-6} 
 & XLM-R\_large & Mod. SLUE & 0.82 & 0.66 & 0.73 \\ \cline{2-6} 
 & XLM-R\_large & seqeval & 0.61 & 0.70 & 0.65 \\ \cline{2-6} 
 & XLM-R\_large & SLUE & 0.80 & 0.64 & 0.71 \\ \hline
\end{tabular}
\caption{NER results of \textbf{ORGANIZATION} entity type (in percent) on \textbf{reference text} of test set using various pre-trained language models. Metrics shown are Precision, Recall, and overall micro/macro F1 score. Evaluation toolkits used are: seqeval, SLUE and our modified SLUE.}
\label{appx_NERresults_ORGANIZATION}
\end{table}
\begin{table}
\centering
\begin{tabular}{|c|l|l|c|c|c|}
\hline
\textbf{Entity Type} & \multicolumn{1}{c|}{\textbf{NER}} & \multicolumn{1}{c|}{\textbf{Eval. Toolkit}} & \textbf{Prec.} & \textbf{Rec.} & \textbf{F1} \\ \hline
\multirow{27}{*}{PERSONALCARE} & BARTpho & Mod. SLUE & 0.17 & 0.77 & 0.28 \\ \cline{2-6} 
 & BARTpho & seqeval & 0.75 & 0.21 & 0.32 \\ \cline{2-6} 
 & BARTpho & SLUE & 0.17 & 0.77 & 0.28 \\ \cline{2-6} 
 & mBART-50 & Mod. SLUE & 0.23 & 0.80 & 0.36 \\ \cline{2-6} 
 & mBART-50 & seqeval & 0.76 & 0.26 & 0.39 \\ \cline{2-6} 
 & mBART-50 & SLUE & 0.23 & 0.80 & 0.36 \\ \cline{2-6} 
 & PhoBERT\_base & Mod. SLUE & 0.18 & 0.87 & 0.30 \\ \cline{2-6} 
 & PhoBERT\_base & seqeval & 0.85 & 0.20 & 0.33 \\ \cline{2-6} 
 & PhoBERT\_base & SLUE & 0.18 & 0.85 & 0.30 \\ \cline{2-6} 
 & PhoBERT\_base-v2 & Mod. SLUE & 0.16 & 0.85 & 0.27 \\ \cline{2-6} 
 & PhoBERT\_base-v2 & seqeval & 0.85 & 0.18 & 0.30 \\ \cline{2-6} 
 & PhoBERT\_base-v2 & SLUE & 0.16 & 0.84 & 0.27 \\ \cline{2-6} 
 & PhoBERT\_large & Mod. SLUE & 0.19 & 0.83 & 0.31 \\ \cline{2-6} 
 & PhoBERT\_large & seqeval & 0.80 & 0.21 & 0.33 \\ \cline{2-6} 
 & PhoBERT\_large & SLUE & 0.19 & 0.82 & 0.31 \\ \cline{2-6} 
 & ViDeBERTa\_base & Mod. SLUE & 0.17 & 0.76 & 0.28 \\ \cline{2-6} 
 & ViDeBERTa\_base & seqeval & 0.68 & 0.14 & 0.23 \\ \cline{2-6} 
 & ViDeBERTa\_base & SLUE & 0.17 & 0.75 & 0.28 \\ \cline{2-6} 
 & ViT5\_base & Mod. SLUE & 0.17 & 0.79 & 0.28 \\ \cline{2-6} 
 & ViT5\_base & seqeval & 0.75 & 0.18 & 0.29 \\ \cline{2-6} 
 & ViT5\_base & SLUE & 0.17 & 0.78 & 0.27 \\ \cline{2-6} 
 & XLM-R\_base & Mod. SLUE & 0.21 & 0.83 & 0.33 \\ \cline{2-6} 
 & XLM-R\_base & seqeval & 0.77 & 0.21 & 0.33 \\ \cline{2-6} 
 & XLM-R\_base & SLUE & 0.20 & 0.81 & 0.32 \\ \cline{2-6} 
 & XLM-R\_large & Mod. SLUE & 0.22 & 0.87 & 0.36 \\ \cline{2-6} 
 & XLM-R\_large & seqeval & 0.84 & 0.24 & 0.38 \\ \cline{2-6} 
 & XLM-R\_large & SLUE & 0.22 & 0.86 & 0.35 \\ \hline
\end{tabular}
\caption{NER results of \textbf{PERSONALCARE} entity type (in percent) on \textbf{reference text} of test set using various pre-trained language models. Metrics shown are Precision, Recall, and overall micro/macro F1 score. Evaluation toolkits used are: seqeval, SLUE and our modified SLUE.}
\label{appx_NERresults_PERSONALCARE}
\end{table}
\begin{table}
\centering
\begin{tabular}{|c|l|l|c|c|c|}
\hline
\textbf{Entity Type} & \multicolumn{1}{c|}{\textbf{NER}} & \multicolumn{1}{c|}{\textbf{Eval. Toolkit}} & \textbf{Prec.} & \textbf{Rec.} & \textbf{F1} \\ \hline
\multirow{27}{*}{PREVENTIVEMED} & BARTpho & Mod. SLUE & 0.01 & 0.15 & 0.01 \\ \cline{2-6} 
 & BARTpho & seqeval & 0.21 & 0.01 & 0.02 \\ \cline{2-6} 
 & BARTpho & SLUE & 0.01 & 0.15 & 0.01 \\ \cline{2-6} 
 & mBART-50 & Mod. SLUE & 0.02 & 0.45 & 0.04 \\ \cline{2-6} 
 & mBART-50 & seqeval & 0.42 & 0.02 & 0.04 \\ \cline{2-6} 
 & mBART-50 & SLUE & 0.02 & 0.45 & 0.04 \\ \cline{2-6} 
 & PhoBERT\_base & Mod. SLUE & 0.03 & 0.75 & 0.06 \\ \cline{2-6} 
 & PhoBERT\_base & seqeval & 0.56 & 0.02 & 0.05 \\ \cline{2-6} 
 & PhoBERT\_base & SLUE & 0.03 & 0.69 & 0.06 \\ \cline{2-6} 
 & PhoBERT\_base-v2 & Mod. SLUE & 0.02 & 0.44 & 0.04 \\ \cline{2-6} 
 & PhoBERT\_base-v2 & seqeval & 0.33 & 0.02 & 0.03 \\ \cline{2-6} 
 & PhoBERT\_base-v2 & SLUE & 0.02 & 0.39 & 0.03 \\ \cline{2-6} 
 & PhoBERT\_large & Mod. SLUE & 0.03 & 0.78 & 0.06 \\ \cline{2-6} 
 & PhoBERT\_large & seqeval & 0.72 & 0.03 & 0.06 \\ \cline{2-6} 
 & PhoBERT\_large & SLUE & 0.03 & 0.75 & 0.06 \\ \cline{2-6} 
 & ViDeBERTa\_base & Mod. SLUE & 0.00 & 0.06 & 0.01 \\ \cline{2-6} 
 & ViDeBERTa\_base & seqeval & 0.00 & 0.00 & 0.00 \\ \cline{2-6} 
 & ViDeBERTa\_base & SLUE & 0.00 & 0.00 & 0.00 \\ \cline{2-6} 
 & ViT5\_base & Mod. SLUE & 0.02 & 0.39 & 0.04 \\ \cline{2-6} 
 & ViT5\_base & seqeval & 0.39 & 0.03 & 0.05 \\ \cline{2-6} 
 & ViT5\_base & SLUE & 0.02 & 0.35 & 0.04 \\ \cline{2-6} 
 & XLM-R\_base & Mod. SLUE & 0.01 & 0.14 & 0.02 \\ \cline{2-6} 
 & XLM-R\_base & seqeval & 0.00 & 0.00 & 0.00 \\ \cline{2-6} 
 & XLM-R\_base & SLUE & 0.01 & 0.08 & 0.01 \\ \cline{2-6} 
 & XLM-R\_large & Mod. SLUE & 0.03 & 0.78 & 0.06 \\ \cline{2-6} 
 & XLM-R\_large & seqeval & 0.72 & 0.03 & 0.05 \\ \cline{2-6} 
 & XLM-R\_large & SLUE & 0.03 & 0.78 & 0.06 \\ \hline
\end{tabular}
\caption{NER results of \textbf{PREVENTIVEMED} entity type (in percent) on \textbf{reference text} of test set using various pre-trained language models. Metrics shown are Precision, Recall, and overall micro/macro F1 score. Evaluation toolkits used are: seqeval, SLUE and our modified SLUE.}
\label{appx_NERresults_PREVENTIVEMED}
\end{table}
\begin{table}
\centering
\begin{tabular}{|c|l|l|c|c|c|}
\hline
\textbf{Entity Type} & \multicolumn{1}{c|}{\textbf{NER}} & \multicolumn{1}{c|}{\textbf{Eval. Toolkit}} & \textbf{Prec.} & \textbf{Rec.} & \textbf{F1} \\ \hline
\multirow{27}{*}{SURGERY} & BARTpho & Mod. SLUE & 0.57 & 0.57 & 0.57 \\ \cline{2-6} 
 & BARTpho & seqeval & 0.48 & 0.55 & 0.51 \\ \cline{2-6} 
 & BARTpho & SLUE & 0.57 & 0.56 & 0.57 \\ \cline{2-6} 
 & mBART-50 & Mod. SLUE & 0.67 & 0.38 & 0.48 \\ \cline{2-6} 
 & mBART-50 & seqeval & 0.29 & 0.55 & 0.38 \\ \cline{2-6} 
 & mBART-50 & SLUE & 0.64 & 0.37 & 0.47 \\ \cline{2-6} 
 & PhoBERT\_base & Mod. SLUE & 0.70 & 0.48 & 0.56 \\ \cline{2-6} 
 & PhoBERT\_base & seqeval & 0.36 & 0.50 & 0.42 \\ \cline{2-6} 
 & PhoBERT\_base & SLUE & 0.68 & 0.46 & 0.55 \\ \cline{2-6} 
 & PhoBERT\_base-v2 & Mod. SLUE & 0.79 & 0.66 & 0.72 \\ \cline{2-6} 
 & PhoBERT\_base-v2 & seqeval & 0.62 & 0.59 & 0.61 \\ \cline{2-6} 
 & PhoBERT\_base-v2 & SLUE & 0.78 & 0.65 & 0.71 \\ \cline{2-6} 
 & PhoBERT\_large & Mod. SLUE & 0.85 & 0.46 & 0.59 \\ \cline{2-6} 
 & PhoBERT\_large & seqeval & 0.38 & 0.66 & 0.49 \\ \cline{2-6} 
 & PhoBERT\_large & SLUE & 0.85 & 0.45 & 0.59 \\ \cline{2-6} 
 & ViDeBERTa\_base & Mod. SLUE & 0.78 & 0.32 & 0.46 \\ \cline{2-6} 
 & ViDeBERTa\_base & seqeval & 0.24 & 0.45 & 0.32 \\ \cline{2-6} 
 & ViDeBERTa\_base & SLUE & 0.76 & 0.31 & 0.44 \\ \cline{2-6} 
 & ViT5\_base & Mod. SLUE & 0.61 & 0.69 & 0.65 \\ \cline{2-6} 
 & ViT5\_base & seqeval & 0.55 & 0.51 & 0.52 \\ \cline{2-6} 
 & ViT5\_base & SLUE & 0.60 & 0.69 & 0.64 \\ \cline{2-6} 
 & XLM-R\_base & Mod. SLUE & 0.66 & 0.67 & 0.66 \\ \cline{2-6} 
 & XLM-R\_base & seqeval & 0.59 & 0.46 & 0.52 \\ \cline{2-6} 
 & XLM-R\_base & SLUE & 0.66 & 0.66 & 0.66 \\ \cline{2-6} 
 & XLM-R\_large & Mod. SLUE & 0.80 & 0.46 & 0.59 \\ \cline{2-6} 
 & XLM-R\_large & seqeval & 0.39 & 0.65 & 0.49 \\ \cline{2-6} 
 & XLM-R\_large & SLUE & 0.80 & 0.46 & 0.58 \\ \hline
\end{tabular}
\caption{NER results of \textbf{SURGERY} entity type (in percent) on \textbf{reference text} of test set using various pre-trained language models. Metrics shown are Precision, Recall, and overall micro/macro F1 score. Evaluation toolkits used are: seqeval, SLUE and our modified SLUE.}
\label{appx_NERresults_SURGERY}
\end{table}
\begin{table}
\centering
\begin{tabular}{|c|l|l|c|c|c|}
\hline
\textbf{Entity Type} & \multicolumn{1}{c|}{\textbf{NER}} & \multicolumn{1}{c|}{\textbf{Eval. Toolkit}} & \textbf{Prec.} & \textbf{Rec.} & \textbf{F1} \\ \hline
\multirow{27}{*}{TRANSPORTATION} & BARTpho & Mod. SLUE & 0.98 & 0.96 & 0.97 \\ \cline{2-6} 
 & BARTpho & seqeval & 0.93 & 0.93 & 0.93 \\ \cline{2-6} 
 & BARTpho & SLUE & 0.98 & 0.96 & 0.97 \\ \cline{2-6} 
 & mBART-50 & Mod. SLUE & 1,00 & 0.70 & 0.82 \\ \cline{2-6} 
 & mBART-50 & seqeval & 0.67 & 0.95 & 0.78 \\ \cline{2-6} 
 & mBART-50 & SLUE & 1,00 & 0.70 & 0.82 \\ \cline{2-6} 
 & PhoBERT\_base & Mod. SLUE & 1,00 & 0.90 & 0.95 \\ \cline{2-6} 
 & PhoBERT\_base & seqeval & 0.85 & 0.92 & 0.88 \\ \cline{2-6} 
 & PhoBERT\_base & SLUE & 1,00 & 0.90 & 0.95 \\ \cline{2-6} 
 & PhoBERT\_base-v2 & Mod. SLUE & 1,00 & 0.69 & 0.82 \\ \cline{2-6} 
 & PhoBERT\_base-v2 & seqeval & 0.59 & 0.80 & 0.68 \\ \cline{2-6} 
 & PhoBERT\_base-v2 & SLUE & 1,00 & 0.69 & 0.82 \\ \cline{2-6} 
 & PhoBERT\_large & Mod. SLUE & 0.95 & 0.91 & 0.93 \\ \cline{2-6} 
 & PhoBERT\_large & seqeval & 0.89 & 0.92 & 0.91 \\ \cline{2-6} 
 & PhoBERT\_large & SLUE & 0.95 & 0.91 & 0.93 \\ \cline{2-6} 
 & ViDeBERTa\_base & Mod. SLUE & 0.00 & 0.00 & 0.00 \\ \cline{2-6} 
 & ViDeBERTa\_base & seqeval & 0.00 & 0.00 & 0.00 \\ \cline{2-6} 
 & ViDeBERTa\_base & SLUE & 0.00 & 0.00 & 0.00 \\ \cline{2-6} 
 & ViT5\_base & Mod. SLUE & 0.00 & 0.00 & 0.00 \\ \cline{2-6} 
 & ViT5\_base & seqeval & 0.00 & 0.00 & 0.00 \\ \cline{2-6} 
 & ViT5\_base & SLUE & 0.00 & 0.00 & 0.00 \\ \cline{2-6} 
 & XLM-R\_base & Mod. SLUE & 0.00 & 0.00 & 0.00 \\ \cline{2-6} 
 & XLM-R\_base & seqeval & 0.00 & 0.00 & 0.00 \\ \cline{2-6} 
 & XLM-R\_base & SLUE & 0.00 & 0.00 & 0.00 \\ \cline{2-6} 
 & XLM-R\_large & Mod. SLUE & 1,00 & 0.72 & 0.84 \\ \cline{2-6} 
 & XLM-R\_large & seqeval & 0.63 & 0.81 & 0.71 \\ \cline{2-6} 
 & XLM-R\_large & SLUE & 1,00 & 0.72 & 0.84 \\ \hline
\end{tabular}
\caption{NER results of \textbf{TRANSPORTATION} entity type (in percent) on \textbf{reference text} of test set using various pre-trained language models. Metrics shown are Precision, Recall, and overall micro/macro F1 score. Evaluation toolkits used are: seqeval, SLUE and our modified SLUE.}
\label{appx_NERresults_TRANSPORTATION}
\end{table}
\begin{table}
\centering
\begin{tabular}{|c|l|l|c|c|c|}
\hline
\textbf{Entity Type} & \multicolumn{1}{c|}{\textbf{NER}} & \multicolumn{1}{c|}{\textbf{Eval. Toolkit}} & \textbf{Prec.} & \textbf{Rec.} & \textbf{F1} \\ \hline
\multirow{27}{*}{TREATMENT} & BARTpho & Mod. SLUE & 0.63 & 0.82 & 0.71 \\ \cline{2-6} 
 & BARTpho & seqeval & 0.75 & 0.57 & 0.65 \\ \cline{2-6} 
 & BARTpho & SLUE & 0.63 & 0.82 & 0.71 \\ \cline{2-6} 
 & mBART-50 & Mod. SLUE & 0.66 & 0.83 & 0.74 \\ \cline{2-6} 
 & mBART-50 & seqeval & 0.81 & 0.61 & 0.70 \\ \cline{2-6} 
 & mBART-50 & SLUE & 0.66 & 0.83 & 0.74 \\ \cline{2-6} 
 & PhoBERT\_base & Mod. SLUE & 0.66 & 0.88 & 0.75 \\ \cline{2-6} 
 & PhoBERT\_base & seqeval & 0.86 & 0.60 & 0.71 \\ \cline{2-6} 
 & PhoBERT\_base & SLUE & 0.66 & 0.88 & 0.75 \\ \cline{2-6} 
 & PhoBERT\_base-v2 & Mod. SLUE & 0.62 & 0.88 & 0.73 \\ \cline{2-6} 
 & PhoBERT\_base-v2 & seqeval & 0.87 & 0.56 & 0.68 \\ \cline{2-6} 
 & PhoBERT\_base-v2 & SLUE & 0.62 & 0.88 & 0.72 \\ \cline{2-6} 
 & PhoBERT\_large & Mod. SLUE & 0.59 & 0.87 & 0.70 \\ \cline{2-6} 
 & PhoBERT\_large & seqeval & 0.86 & 0.53 & 0.65 \\ \cline{2-6} 
 & PhoBERT\_large & SLUE & 0.59 & 0.87 & 0.70 \\ \cline{2-6} 
 & ViDeBERTa\_base & Mod. SLUE & 0.62 & 0.86 & 0.72 \\ \cline{2-6} 
 & ViDeBERTa\_base & seqeval & 0.84 & 0.52 & 0.64 \\ \cline{2-6} 
 & ViDeBERTa\_base & SLUE & 0.62 & 0.86 & 0.72 \\ \cline{2-6} 
 & ViT5\_base & Mod. SLUE & 0.46 & 0.84 & 0.60 \\ \cline{2-6} 
 & ViT5\_base & seqeval & 0.81 & 0.41 & 0.55 \\ \cline{2-6} 
 & ViT5\_base & SLUE & 0.46 & 0.84 & 0.60 \\ \cline{2-6} 
 & XLM-R\_base & Mod. SLUE & 0.48 & 0.86 & 0.62 \\ \cline{2-6} 
 & XLM-R\_base & seqeval & 0.84 & 0.40 & 0.54 \\ \cline{2-6} 
 & XLM-R\_base & SLUE & 0.48 & 0.86 & 0.62 \\ \cline{2-6} 
 & XLM-R\_large & Mod. SLUE & 0.59 & 0.89 & 0.71 \\ \cline{2-6} 
 & XLM-R\_large & seqeval & 0.89 & 0.54 & 0.67 \\ \cline{2-6} 
 & XLM-R\_large & SLUE & 0.59 & 0.89 & 0.71 \\ \hline
\end{tabular}
\caption{NER results of \textbf{TREATMENT} entity type (in percent) on \textbf{reference text} of test set using various pre-trained language models. Metrics shown are Precision, Recall, and overall micro/macro F1 score. Evaluation toolkits used are: seqeval, SLUE and our modified SLUE.}
\label{appx_NERresults_TREATMENT}
\end{table}
\begin{table}
\centering
\begin{tabular}{|c|l|l|c|c|c|}
\hline
\textbf{Entity Type} & \multicolumn{1}{c|}{\textbf{NER}} & \multicolumn{1}{c|}{\textbf{Eval. Toolkit}} & \textbf{Prec.} & \textbf{Rec.} & \textbf{F1} \\ \hline
\multirow{27}{*}{UNITCALIBRATOR} & BARTpho & Mod. SLUE & 0.42 & 0.65 & 0.51 \\ \cline{2-6} 
 & BARTpho & seqeval & 0.52 & 0.31 & 0.39 \\ \cline{2-6} 
 & BARTpho & SLUE & 0.41 & 0.63 & 0.50 \\ \cline{2-6} 
 & mBART-50 & Mod. SLUE & 0.38 & 0.58 & 0.46 \\ \cline{2-6} 
 & mBART-50 & seqeval & 0.45 & 0.26 & 0.33 \\ \cline{2-6} 
 & mBART-50 & SLUE & 0.37 & 0.56 & 0.44 \\ \cline{2-6} 
 & PhoBERT\_base & Mod. SLUE & 0.44 & 0.73 & 0.55 \\ \cline{2-6} 
 & PhoBERT\_base & seqeval & 0.61 & 0.30 & 0.41 \\ \cline{2-6} 
 & PhoBERT\_base & SLUE & 0.43 & 0.71 & 0.53 \\ \cline{2-6} 
 & PhoBERT\_base-v2 & Mod. SLUE & 0.46 & 0.74 & 0.57 \\ \cline{2-6} 
 & PhoBERT\_base-v2 & seqeval & 0.61 & 0.33 & 0.43 \\ \cline{2-6} 
 & PhoBERT\_base-v2 & SLUE & 0.45 & 0.72 & 0.55 \\ \cline{2-6} 
 & PhoBERT\_large & Mod. SLUE & 0.48 & 0.73 & 0.58 \\ \cline{2-6} 
 & PhoBERT\_large & seqeval & 0.60 & 0.34 & 0.43 \\ \cline{2-6} 
 & PhoBERT\_large & SLUE & 0.47 & 0.71 & 0.56 \\ \cline{2-6} 
 & ViDeBERTa\_base & Mod. SLUE & 0.32 & 0.44 & 0.37 \\ \cline{2-6} 
 & ViDeBERTa\_base & seqeval & 0.20 & 0.10 & 0.14 \\ \cline{2-6} 
 & ViDeBERTa\_base & SLUE & 0.28 & 0.39 & 0.33 \\ \cline{2-6} 
 & ViT5\_base & Mod. SLUE & 0.34 & 0.63 & 0.44 \\ \cline{2-6} 
 & ViT5\_base & seqeval & 0.50 & 0.24 & 0.32 \\ \cline{2-6} 
 & ViT5\_base & SLUE & 0.33 & 0.62 & 0.43 \\ \cline{2-6} 
 & XLM-R\_base & Mod. SLUE & 0.44 & 0.72 & 0.54 \\ \cline{2-6} 
 & XLM-R\_base & seqeval & 0.54 & 0.27 & 0.36 \\ \cline{2-6} 
 & XLM-R\_base & SLUE & 0.42 & 0.69 & 0.52 \\ \cline{2-6} 
 & XLM-R\_large & Mod. SLUE & 0.50 & 0.75 & 0.60 \\ \cline{2-6} 
 & XLM-R\_large & seqeval & 0.65 & 0.37 & 0.47 \\ \cline{2-6} 
 & XLM-R\_large & SLUE & 0.49 & 0.74 & 0.59 \\ \hline
\end{tabular}
\caption{NER results of \textbf{UNITCALIBRATOR} entity type (in percent) on \textbf{reference text} of test set using various pre-trained language models. Metrics shown are Precision, Recall, and overall micro/macro F1 score. Evaluation toolkits used are: seqeval, SLUE and our modified SLUE.}
\label{appx_NERresults_UNITCALIBRATOR}
\end{table}
\input{tabs_and_figs/appx_NERresults_OverallMacro}
\input{tabs_and_figs/appx_NERresults_OverallMicro}

\begin{table}
\centering
\begin{tabular}{|c|l|l|l|c|c|c|}
\hline
\textbf{Entity Type} & \multicolumn{1}{c|}{\textbf{NER}} & \multicolumn{1}{c|}{\textbf{ASR}} & \multicolumn{1}{c|}{\textbf{Eval. Toolkit}} & \multicolumn{1}{c|}{\textbf{Prec.}} & \multicolumn{1}{c|}{\textbf{Rec.}} & \multicolumn{1}{c|}{\textbf{F1}} \\ \hline
\multirow{36}{*}{AGE} & ViDeBERTa\_base & XLSR-53-Viet & SLUE & 0.24 & 0.03 & 0.05 \\ \cline{2-7} 
 & ViDeBERTa\_base & XLSR-53-Viet & Mod. SLUE & 0.32 & 0.04 & 0.07 \\ \cline{2-7} 
 & ViDeBERTa\_base & w2v2-Viet & SLUE & 0.24 & 0.03 & 0.05 \\ \cline{2-7} 
 & ViDeBERTa\_base & w2v2-Viet & Mod. SLUE & 0.33 & 0.04 & 0.07 \\ \cline{2-7} 
 & ViT5\_base & XLSR-53-Viet & SLUE & 0.63 & 0.49 & 0.55 \\ \cline{2-7} 
 & ViT5\_base & XLSR-53-Viet & Mod. SLUE & 0.68 & 0.53 & 0.59 \\ \cline{2-7} 
 & ViT5\_base & w2v2-Viet & SLUE & 0.63 & 0.52 & 0.57 \\
\hline
\end{tabular}
\caption{NER results of \textbf{AGE} entity type (in percent) on \textbf{ASR output} of test set using various pre-trained language models and ASR models. Metrics shown are Precision, Recall, and overall micro/macro F1 score. Evaluation toolkits used are: SLUE and our modified SLUE.}
\label{appx_spk_NERresults_AGE}
\end{table}
\input{tabs_and_figs/appx_spk_NERresults_DATETIME}
\input{tabs_and_figs/appx_spk_NERresults_DIAGNOSTICS}
\input{tabs_and_figs/appx_spk_NERresults_DISEASESYMTOM}
\input{tabs_and_figs/appx_spk_NERresults_DRUGCHEMICAL}
\input{tabs_and_figs/appx_spk_NERresults_FOODDRINK}
\input{tabs_and_figs/appx_spk_NERresults_GENDER}
\input{tabs_and_figs/appx_spk_NERresults_LOCATION}
\input{tabs_and_figs/appx_spk_NERresults_MEDDEVICETECHNIQUE}
\input{tabs_and_figs/appx_spk_NERresults_OCCUPATION}
\input{tabs_and_figs/appx_spk_NERresults_ORGAN}
\input{tabs_and_figs/appx_spk_NERresults_ORGANIZATION}
\input{tabs_and_figs/appx_spk_NERresults_PERSONALCARE}
\input{tabs_and_figs/appx_spk_NERresults_PREVENTIVEMED}
\input{tabs_and_figs/appx_spk_NERresults_SURGERY}
\input{tabs_and_figs/appx_spk_NERresults_TRANSPORTATION}
\input{tabs_and_figs/appx_spk_NERresults_TREATMENT}
\input{tabs_and_figs/appx_spk_NERresults_UNITCALIBRATOR}
\input{tabs_and_figs/appx_spk_NERresults_OverallMacro}
\input{tabs_and_figs/appx_spk_NERresults_OverallMicro}
\end{document}